\documentstyle[donrack,rac112,epsfig]{rackham}
\input{epsf} 
\begin{document}
\newcounter{old1}
\newcounter{old2}
\newcounter{old3}
\newcounter{old4}
\titlepage{%
RF depolarizing resonances in the presence of a full Siberian snake and full snake spin-flipping}
{{\bf Boris B. Blinov}}
{Doctor of Philosophy}
{Physics}
{2000}
{  Professor Alan D.~Krisch, Chair\\
   Professor Ratindranath Akhoury\\
   Professor Ronald M. Gilgenbach\\
   Associate Research Scientist Ali M.T. Lin\\
   Professor Rudolf P. Thun\\
}
\clearpage
\startabstractpage{RF depolarizing resonances in the presence of a full Siberian snake and 
full snake spin-flipping}
{Boris B. Blinov}
{Alan D. Krisch}
Frequent polarization reversals, or spin-flips, of a stored polarized beam in a high energy scattering
asymmetry experiments may greatly reduce systematic errors of spin asymmetry
measurements. 
A spin-flipping technique is being developed by using rf magnets running at a frequency
close to the spin precession frequency, thereby creating spin-depolarizing resonances;
the spin can then be flipped by ramping the rf magnet's frequency through the resonance.
We studied, at the Indiana University Cyclotron Facility Cooler Ring, properties of such rf 
depolarizing resonances in the presence of a nearly-full Siberian snake and their possible
application for spin-flipping. By using an rf-solenoid magnet, we reached a 98.7$\pm$1$\%$
efficiency of spin-flipping. However, an rf-dipole magnet is more practical at high energies;
hence, studies of spin-flipping by an rf-dipole are underway at IUCF.
\clearpage
\pagestyle{chappage}
\pagenumbering{roman}
\setcounter{page}{2}
\dedicationpage{For my wife}
\startacknowledgementspage
I would like to thank Professor Alan Krisch for his valuable advice,
enouragement and support. I also thank my dissertation
committee members their helpful comments and suggestions to
this thesis. I thank my fellow
collaborators from Michigan, Indiana, Protvino and Brookhaven for many
hours they devoted to this experiment.
I would especially like to thank Rick Phelps for teaching me
a lot about how the experiment works.

I would like to thank the IUCF technical staff for their hard work on
successful experiment operation. In particular, I want to thank the operators
headed by Gary East with help from Terry Sloan for giving us beam when we
needed it, 
the cryogenics experts Kevin Komisarcik and John Vanderwerp for keeping
our snake cool, and the polarizaed ion source group headed by Vladimir Derenchuk
for making polarized protons out of hydrogen gas.

Finally, I would like to thank my parents, my big brother and my
little sister for their love and support, and my wife Svetlana for always being there
when I needed her.
\clearpage
\tableofcontents
\clearpage
\addcontentsline{toc}{chapter}{LIST OF TABLES}
\listoftables
\clearpage
\addcontentsline{toc}{chapter}{LIST OF FIGURES} 
\listoffigures
\clearpage
\pagenumbering{arabic}
\addtocontents{toc}{\vskip 0.5cm\relax }

\chapter{Introduction}

Spin is one of the fundamental quantum numbers. Studying effects that involve
spin is crucial for better understanding Nature. Such studies can be 
done using polarized beams, polarized targets, or both. While polarized proton
targets are now widely used in a variety of experiments, when combined with a
polarized proton beam they can provide much more information. 
Thus, developing a high-energy polarized proton beam for physics experiments is
very useful; it is also very challenging.

In a circular accelerator, each proton's spin precesses around the vertical
magnetic fields of the accelerator's ring's bending dipoles. At each 
point of the orbit,
a vector called the {\bf stable spin direction} (SSD) is defined such that the
spin component along this direction remains unchanged from turn to turn. 
Any spin component,
which is not aligned with the SSD, will precess around it, with a frequency 
called the {\bf spin precession frequency} $f_s$ that is
related to the proton's circulation frequency $f_c$ by:
\begin{equation}
f_s = f_c \nu_s,
\end{equation}
where $\nu_s$ is the {\bf spin tune}, which is the number of spin precessions 
during one turn around the ring.

If there are only vertical magnetic fields, then the polarization remains
unchanged; however, any horizontal magnetic field can destroy the beam 
polarization. A {\bf depolarizing resonance} occurs whenever there 
is a periodic horizontal
magnetic field whose frequency is equal to the spin precession
frequency; in terms of the tunes, this occurs when
\begin{equation}
\nu_s=n \nu_{d}+k,
\end{equation}
where $\nu_{d}$ is the tune of the depolarizing magnetic field, 
and $n$ and $k$ are 
integers. The spin tune $\nu_s$ is proportional to the proton's 
energy:
\begin{equation}
\nu_s = G \gamma,
\end{equation}
where G=1.792847 is the proton's anomalous magnetic moment, and $\gamma$ is
the Lorentz energy factor. Since a resonance occurs at least every 523~MeV, 
the protons will encounter many depolarizing
resonances during acceleration to a high energy.

Three main types of depolarizing resonances exist in a circular accelerator ring.
The first type, which is called an {\bf imperfection resonance}, 
is caused by the
imperfections in the ring's magnetic structure. An imperfection depolarizing
resonance occurs whenever the spin tune is equal to an integer
\begin{equation}
\nu_s = n.
\end{equation}

The second type of depolarizing resonances is called an {\bf intrinsic
resonance}. These resonances are caused by the horizontal magnetic fields
in the ring's quadrupoles, which are an intrinsic property of any 
strong focusing ring. 
The intrinsic depolarizing resonance
condition is:
\begin{equation}
\nu_s = l \nu_x + m \nu_y + n,
\end{equation}
where $\nu_x$ and $\nu_y$ are the horizontal and vertical betatron 
tunes, while $l$, $m$ and $n$ are integers. The $\nu_y$ resonances are normally
much stronger because the vertical oscillations move the protons into the horizontal
fields of the quadrupoles.

The synchrotron oscillations of the protons' energy cause the third type of
depolarizing resonance, which is called a {\bf synchrotron depolarizing
resonance}; its resonance condition is:
\begin{equation}
\nu_s = k \nu_{sync} + n,
\end{equation}
where $\nu_{sync}$ is the synchrotron frequency, while $k$ and $n$ are
integers.

Many experimental techniques have been proposed to overcome these depolarizing 
resonances. Some techniques involve complicated betatron tune manipulations
to quickly ``jump'' through a resonance, or using special correction dipoles
to overcome some other resonances~\cite{zgs,ags,saturne,kek}. The main disadvantage of these methods
is that each resonance must be corrected individually, which becomes
impractical at very high energies, where there are many depolarizing resonances.

A more recent method, proposed by Derbenev and Kondratenko in 1978~\cite{snake}, 
involves using
a spin rotator called a {\bf Siberian snake} that makes the spin tune 
equal to 1/2 independent of the beam energy. This energy-independent $\nu_s$ eliminates most of the
depolarizing resonances.

Once a polarized proton beam is accelerated to a high energy and stored, it is 
important to be able to reverse its polarization direction 
in order to reduce the systematic error in the
asymmetry, measured in various polarized scattering experiments.

Studies at the IUCF Cooler Ring show that these polarization reversals
(spin-flips) can be successfully done using an rf magnet. An rf magnet
creates an {\bf rf depolarizing resonance}, whose
resonance condition is:
\begin{equation}
f_s = f_{rf} + n f_c,
\end{equation}
where $f_{rf}$ is the frequency of the rf magnet, $f_c$ is the protons'
circulation frequency, and $n$ is an integer. By varying the 
rf magnet's frequency from a value below the depolarizing resonance to a value 
above the resonance, one
can {\bf cross} the resonance. After such a crossing, the final beam polarization 
$P_f$ is related to the initial beam polarization $P_i$ via the 
{\bf Froissart-Stora formula}~\cite{stora}:
\begin{equation}
P_f = P_i[2 e^{- \frac{\mbox{\scriptsize $(\pi \epsilon f_c)^2$}}{\mbox{\scriptsize $(\Delta f / \Delta t)$}}} - 1],
\end{equation}
where $\epsilon$ is the resonance strength, 
and $\Delta f / \Delta t$ is the
resonance crossing rate, while $\Delta f$ is the frequency's range during its linear
variation time $\Delta t$. The value of the exponent determines
the final polarization. If the resonance strength is small
and/or the crossing rate is fast, then the final polarization is very close to
the initial polarization; this is called {\bf fast crossing}. For larger
resonance strengths and slower crossing rates, the exponent becomes larger, so
that the exponential becomes smaller; then the final polarization can decrease
towards zero; this is called {\bf depolarization}. If the resonance
is very strong and/or the crossing rate is very slow, then the value of the
exponent is very large and the exponential approaches zero; 
therefore, the final polarization is reversed
with respect to the initial polarization, while its absolute value is the
same; this is called {\bf spin flip}.

This thesis describes
a spin-flipping experiment in the presence of a nearly-full Siberian
snake at the IUCF Cooler Ring.
Chapter II contains a brief discussion of polarized beam theory.
I do not try to cover the whole field of polarized beam theory;
instead I stress rf-induced depolarizing resonances and
spin-flipping. Chapter III describes the experimental apparatus. 
In Chapter IV, the experimental results are discussed, while Chapter V
contains the conclusions.

\chapter{THEORETICAL MOTIVATION}

\section{Introduction}

In this chapter we first discuss spin motion in the magnetic fields of a circular
accelerator ring. The effects
of a full and nearly full Siberian snake on the spin tune and the stable
spin direction are then described. Later we discuss rf-induced 
resonances and
their strength; we finish with discussions of the Froissart-Stora formula 
and spin-flipping by an rf-induced resonance .

\section{Spin Motion in a Magnetic Field}

The Thomas-BMT equation~\cite{thomas,bmt} describes the behavior of a proton's 
spin polarization vector $\vec{S}$ in a magnetic field $\vec{B}$ and an electric field $\vec{E}$: 
\begin{equation}
\frac{d \vec{S}}{dt} = \frac{e}{m_p \gamma}\vec{S} \times [(1 + G \gamma)\vec{B}_{\perp} + (1 + G)\vec{B}_{\parallel} - (G + \frac{1}{1 + \gamma})\gamma \vec{\beta}\times \frac{\vec{E}}{c}],
\end{equation}
where $m_p$ is the proton's rest mass and e is its charge,
while $\vec{B}_{\perp}$ and
$\vec{B}_{\parallel}$ are the magnetic field components transverse and parallel to the 
proton's velocity $\vec{v}$, and $\vec{\beta} = \vec{v}/c$. 
The electric and magnetic fields are measured in the laboratory frame,
while the spin is defined in the proton's rest frame. The electric field term
plays a very minor role in the spin motion in a high energy circular accelerator; thus it can
be neglected.

We shall use the curvilinear coordinate system, which rotates as one moves
around the ring along a proton's ideal orbit; then the position
is described as
\begin{equation}
\vec{r}(x,l,y) = \vec{r}_0(l) + x\hat{x} + y\hat{y},
\end{equation}
where vector $\vec{r}_0$ defines a point on the ideal reference orbit which moves
with an ideal proton, the unit vector $\hat{x}$ points radially outward, 
and $\hat{y}$ points vertically up. The third unit vector is defined as 
$\hat{l}$ = $\frac{d\vec{r}_0}{dl}$, so that it points along the longitudinal velocity of
the ideal proton.

Let us first consider the spin motion in the ring bending dipoles' vertical magnetic field:
\begin{equation}
\vec{B}_{\perp} = -B\hat{y}.
\end{equation}
Since $\vec{B}_{\parallel}$ and $\vec{E}$ are both negligible, Equation~II.1 becomes:
\begin{equation}
\frac{d \vec{S}}{dt} = \frac{e}{m_p \gamma}\vec{S} \times [(1 + G \gamma)\vec{B}_{\perp}].
\end{equation}
Inserting Equation~II.11 we then obtain:
\begin{equation}
\frac{d \vec{S}}{dt} = \frac{eB}{m_p \gamma}(1 + G \gamma)(-S_l\hat{x} + S_x\hat{l}).
\end{equation}
Expanding and comparing the $\hat{x}$, $\hat{l}$ and
$\hat{y}$ components, we get:
\begin{equation}
\frac{dS_x}{dt} = - \frac{eB}{m_p \gamma}(1 + G \gamma)S_l,
\end{equation}
\begin{equation}
\hspace{-4 mm}
\frac{dS_l}{dt} = \frac{eB}{m_p \gamma}(1 + G \gamma)S_x,
\end{equation}
\begin{equation}
\hspace{-31 mm}
\frac{dS_y}{dt} = 0.
\end{equation}
The solution for this system of differential equations is given by
\begin{equation}
S_x = S_{h}cos\left(\frac{eB}{m_p \gamma}(1 + G \gamma)t\right),
\end{equation}
\begin{equation}
\hspace{-3 mm}
S_l = S_{h}sin\left(\frac{eB}{m_p \gamma}(1 + G \gamma)t\right),
\end{equation}
\begin{equation}
\hspace{-36 mm}
S_y = S_{v},
\end{equation}
where $S_{h}$ is the horizontal component of the spin-polarization vector, 
and $S_{v}$ is its vertical component; note that $S_{h}^2 + S_{v}^2=1$.
Equations~II.17-II.19 define the precession of the spin-polarization vector $\vec{S}$ around 
the vertical axis
with an angular velocity 
\begin{equation}
\omega_v = \frac{eB}{m_p \gamma}(1 + G \gamma). 
\end{equation}

Thus, when a proton passes through a dipole field of strength $\int{\!B\!\cdot\!dl}$, 
its spin-polarization rotates by an angle
\begin{equation}
\theta_S = \frac{e\int{\!B\!\cdot\!dl}(1 + G\gamma)}{\gamma m_pv}.
\end{equation}
Notice that, inside this dipole field, the proton's orbit bends by an angle
\begin{equation}
\theta_B = \frac{e\int{\!B\!\cdot\!dl}}{\gamma m_pv};
\end{equation}
therefore, the spin precession angle in the proton's rest frame is:
\begin{equation}
\theta_{sp} = \frac{Ge\int{\!B\!\cdot\!dl}}{m_pv}.
\end{equation}
It can similarly be shown that a proton's spin would precess by this 
angle in any
transverse magnetic field of strength $\int{\!B\!\cdot\!dl}$. Notice 
that this spin precession angle relates quite simply to the proton's bend angle, 
defined by Equation~II.22,  via
\begin{equation}
\theta_{sp} = G\gamma\theta_{B}.
\end{equation}
We can then relate the spin-polarization vector $\vec{S}'$ after passing through 
a dipole to the initial
spin vector $\vec{S}$ by the matrix equation:
\begin{equation}
\vec{S}' = M(\theta)\vec{S},
\end{equation}
where the dipole's precession matrix is given by:
\begin{equation}
M(\theta) = \left (\begin{array}{ccc}
		cos(G\gamma\theta_B) & -sin(G\gamma\theta_B) & 0 \\
		sin(G\gamma\theta_B) & cos(G\gamma\theta_B) & 0 \\
		0 & 0 &1
		\end{array} \right).
\end{equation}

Now, consider a solenoidal field:
\begin{equation}
\vec{B}_{\parallel} = B\hat{l}.
\end{equation}
Since $\vec{B}_{\perp}$ and $\vec{E}$ are now negligible, the Equation~II.1 simplifies to:
\begin{equation}
\frac{d \vec{S}}{dt} = \frac{e}{m_p \gamma}\vec{S} \times [(1 + G)\vec{B}_{\parallel}].
\end{equation}
Inserting Equation~II.27, we obtain:
\begin{equation}
\frac{d \vec{S}}{dt} = \frac{eB}{m_p \gamma}(1 + G)(-S_y\hat{x} + S_x\hat{y}).
\end{equation}
Again, expanding and comparing 
the $\hat{x}$, $\hat{l}$ and $\hat{y}$ components, we get:
\begin{equation}
\frac{dS_x}{dt} = - \frac{eB}{m_p \gamma}(1 + G)S_y,
\end{equation}
\begin{equation}
\hspace{-30 mm}
\frac{dS_l}{dt} = 0,
\end{equation}
\begin{equation}
\hspace{-4 mm}
\frac{dS_y}{dt} = \frac{eB}{m_p \gamma}(1 + G)S_x.
\end{equation}
Similarly, the solution is given by
\begin{equation}
S_x = S_{\perp}cos\left(\frac{eB}{m_p \gamma}(1 + G)t\right),
\end{equation}
\begin{equation}
\hspace{-22mm}
S_l = constant,
\end{equation}
\begin{equation}
S_y = S_{\perp}sin\left(\frac{eB}{m_p \gamma}(1 + G)t\right),
\end{equation}
where $S_{\perp}=\sqrt{S_x^2+S_y^2}$ is the transverse component of the spin-polarization vector 
$\vec{S}$ and $S_l$ is its longitudinal component; note that
$S_{\perp}^2 + S_l^2 = 1$.
As defined by Equations~II.33-II.35, the vector $\vec{S}$ precesses
around the longitudinal axis with an angular velocity 
\begin{equation}
\omega_l~=~\frac{eB}{m_p \gamma}(1 + G). 
\end{equation}
Using Equation~II.36, one can show that when a proton with velocity $\vec{v}$ 
passes through a solenoid field of strength $\int{\!B\!\cdot\!dl}$, its spin rotates by an angle
\begin{equation}
\theta_{sp} = \frac{(1 + G)e\int{\!B\!\cdot\!dl}}{\gamma m_pv}
\end{equation}

We shall define the {\bf snake strength} $s$ of a solenoid magnet as
\begin{equation}
s = \frac{\theta_{sp}}{\pi}.
\end{equation}
Combining Equation~II.37 and Equation~II.38, we can express the solenoidal
field integral, required for a snake strength of $s$:
\begin{equation}
\int{\!B\!\cdot\!dl} = \frac{\pi s\gamma m_pv}{(1 + G)e} = 3.7521\cdot\!s\!\cdot\!p,
\end{equation}
where $p$ is the proton's momentum  in GeV/c.
Similarly to the dipole case, we can define the spin precession 
matrix inside a solenoid:
\begin{equation}
M(s) = \left (\begin{array}{ccc}
		cos(\pi s) & 0 & -sin(\pi s) \\
		0 & 1 & 0 \\
		sin(\pi s) & 0 &cos(\pi s)
		\end{array} \right).
\end{equation}

\section{Spin Tune and Stable Spin Direction}

As shown in the previous section, the spin motion in a magnetic structure
can be viewed as a precession around the magnetic field's direction. Thus, 
after one turn around a ring, the spin-polarization
vector's direction at any point in the ring will usually differ 
from the vector's initial direction by some angle. This change in the spin-polarization 
vector's orientation can
be described as a rotation around an axis, called the {\bf stable spin direction SSD}, by
the single turn precession angle $\theta_{st}$, which is proportional to the spin tune $\nu_s$:
\begin{equation}
\theta_{st} = 2\pi\nu_s.
\end{equation}

In discussing the spin tune and the SSD, it is convenient 
to use the Pauli's spinor notation, where the spin is represented by a two-component
complex spinor $\chi$. The expectation value $<S_i>$ of each component of the 
spin-polarization vector is:
\begin{equation}
<S_i> = \chi^{\dagger}\sigma_i\chi,
\end{equation}
where the $\sigma_i$'s are the Pauli spin matrices. A rotation of the spin by an angle $\theta$
around an axis $\hat{n}$ is then given by the transformation of the spinor $\chi$ into a
spinor $\chi '$, defined by:
\begin{equation}
\chi ' = exp\left(\frac{i\theta\vec{\sigma}\cdot\hat{n}}{2}\right)\chi\equiv M\chi,
\end{equation}
where M is the precession matrix. As a proton goes around a ring, it 
encounters many magnets, each with some spin precession matrix.
Note that  this matrix is a 2$\times$2 matrix since it describes a transformation
of the two-component spinor $\chi$.
The spinor $\chi$ after passing through $n$ of these magnets will be transformed 
into a new spinor $\chi^n$ according to the equation:
\begin{equation}
\chi^n = M_nM_{n-1}...M_1\chi,
\end{equation}
where $M_i$ is the precession matrix in the $i^{th}$ magnet; the product of these 2$\times$2
matrices is also a 2$\times$2 matrix.

After completing one turn around a ring, consisting of $n$ magnets, 
a proton will have encountered all the magnetic fields once;
the product of these $n$ precession matrices is the ring's single-turn matrix $M_{st}$:
\begin{equation}
M_{st} = M_nM_{n-1}...M_1 = exp\left(\frac{i2\pi\nu_s\vec{\sigma}\cdot\hat{n}^{SSD}}{2}\right),
\end{equation}
where $\hat{n}^{SSD}$ is the unit vector of the stable spin direction
(SSD), while $2\pi\nu_s$ is the angle of net spin rotation after one turn. 
It can be shown~\cite{tune} that the spin tune is given by the trace of $M_{st}$:
\begin{equation}
\nu_s = \frac{1}{\pi}cos^{-1}\left(\frac{Tr(M_{st})}{2}\right),
\end{equation}
and the components of the stable spin direction are given by:
\begin{equation}
n^{SSD}_i = \frac{1}{2i}\frac{Tr(\sigma_iM_{st})}{sin(\pi\nu_s)}.
\end{equation}

Now we calculate the spin tune and the stable spin direction for two simple cases: 

$a)$ an ideal ring with only the vertical magnetic fields of its bending dipoles, 

$b)$ the same ideal ring with a full solenoidal snake installed. 

\noindent In case $a$, with the dipole field pointing down, the spin rotates only around the 
vertical axis $-\hat{y}$; thus the SSD is vertical everywhere. 
The total spin precession angle is the sum of individual rotations around the vertical 
axis in each ring's dipole. Since the total bending angle in one turn is $2\pi$,
the single-turn spin precession angle is given by Equation~II.24 to be:
\begin{equation}
\theta_{st} = 2\pi G\gamma.
\end{equation}
Comparing this to the Equation~II.41, we find that for this ideal ring with
only vertical field dipoles:
\begin{equation}
\nu_s = G\gamma.
\end{equation}

Now, to this ideal dipole ring we add a full Siberian snake solenoid magnet, having a
snake strength $s =1$. We consider a point on the ring orbit at an angle $\theta$
from the snake, which is at $\theta = 0$. 
The ring's single turn spin precession matrix is then the matrix product of:
a dipole matrix with a vertical precession angle $G\gamma\theta$,
a solenoid matrix with a longitudinal precession angle $\pi$, and another dipole 
matrix with a vertical precession angle $G\gamma(2\pi -\theta)$.
Writing these matrices in the exponential form, we get:
\begin{equation}
M_{st}(\theta) = e^{-\frac{i}{2}\sigma_yG\gamma\theta}e^{-\frac{i}{2}\sigma_l\pi}e^{-\frac{i}{2}\sigma_yG\gamma(2\pi - \theta)}.
\end{equation}
Multiplying the exponentials and using the Pauli matrices relations, 
it can be shown~\cite{tune} that:
\begin{equation}
Tr(M_{st}) = cos(\pi G\gamma)cos(\pi/2) = 0.
\end{equation}
Then from Equation~II.46 the spin tune is given by:
\begin{equation}
\nu_s = \frac{1}{\pi}cos^{-1}\left(\frac{1}{2}Tr(M_{st})\right) = 1/2;
\end{equation}
thus, $\nu_s$ is indeed independent of the proton's energy, as was mentioned in Chapter~I.
For an arbitrary snake strength $s$, one can show~\cite{tune} that the spin tune is:
\begin{equation}
\nu_s = \frac{1}{\pi}cos^{-1}\left(cos(\pi\!G\gamma)cos(\frac{\pi\!s}{2})\right).
\end{equation}

Using Equations~II.47~and~II.50, we can also obtain the
normalized components of the SSD:
\begin{equation}
n_x = sin(G\gamma(\theta - \pi)),
\end{equation}
\begin{equation}
\hspace{2 mm}
n_l = -cos(G\gamma(\theta - \pi)),
\end{equation}
\begin{equation}
\hspace{-24 mm}
n_y = 0.
\end{equation}
Thus, the stable spin direction is always {\bf horizontal} when
one full Siberian snake is present, while the radial and the longitudinal components
change as the SSD rotates by an angle $G\gamma$ during each turn around the ring.

\section{RF Depolarizing Resonances}

An rf magnetic field from either an rf solenoid or an rf dipole magnet can
depolarize the beam. This depolarization occurs when the frequency of the
rf magnet satisfies the resonance condition, given by Equation~I.7. 

Consider a proton stored in an accelerator ring. Assume that its spin tune
is far away from any imperfection or intrinsic resonance condition; the spin
of the proton then precesses unperturbed around the vertical fields of the 
bending dipoles at a frequency $\omega_s$, as measured in the laboratory frame. 
Now introduce an rf field of amplitude $B$, direction $\hat{i}$ and frequency 
$\omega_{rf}$:
\begin{equation}
\vec{B}_{rf} = Bcos(\omega_{rf}t)\hat{i}.
\end{equation}
Now assume that $\hat{i} = \hat{l}$, so that the rf field is
longitudinal; the following discussion can also be applied to the case
of a transverse field. This longitudinal rf magnetic field can be decomposed into
two constant-magnitude counter-rotating fields $\vec{B}_+$ and $\vec{B}_-$ of amplitude
$\frac{1}{2}B$ and frequency $\omega_{rf}$:
\begin{equation}
\vec{B}_+ = \frac{B}{2}(cos(\omega_{rf}t)\hat{l} + sin(\omega_{rf}t)\hat{x}),
\end{equation}
\begin{equation}
\vec{B}_- = \frac{B}{2}(cos(\omega_{rf}t)\hat{l} - sin(\omega_{rf}t)\hat{x}).
\end{equation}
The field $\vec{B}_+$ rotates in the same direction as the spin vector; the
field $\vec{B}_-$ rotates in the opposite direction, as shown in Figure~II.1.
\begin{figure}
\begin{center}
\epsfig{file=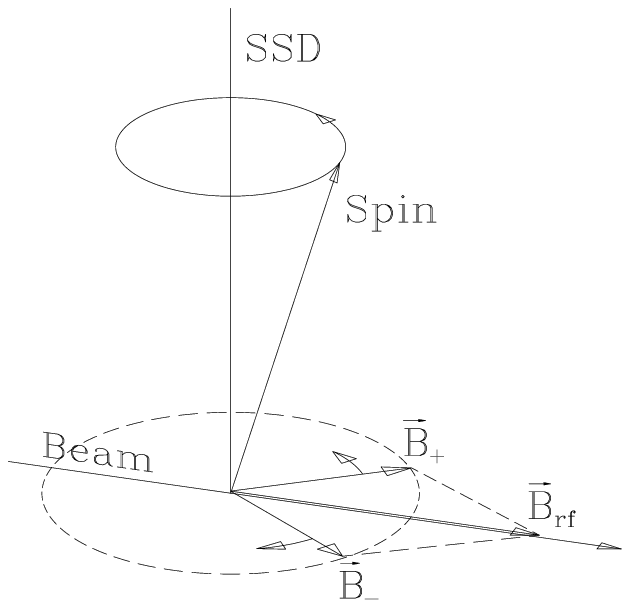}
\end{center}
\caption{\small A longitudinal rf magnetic field decomposition into two counter-rotating fields}
\end{figure}
If the frequency $\omega_{rf}$ is far away from the spin precession
frequency $\omega_s$, the total effect of these fields on the spin 
averages to zero. However, if $\omega_{rf}$ is close to $\omega_s$,
the rotation of the spin vector becomes correlated with the rotation of 
$\vec{B}_+$, which can tilt the stable spin direction. To quantify
this tilt, consider the frame that rotates around the stable spin direction with the
spin precession frequency $\omega_s$. 
As shown in Figure~II.2, in this frame the stable spin direction would be 
\begin{figure}
\epsfig{file=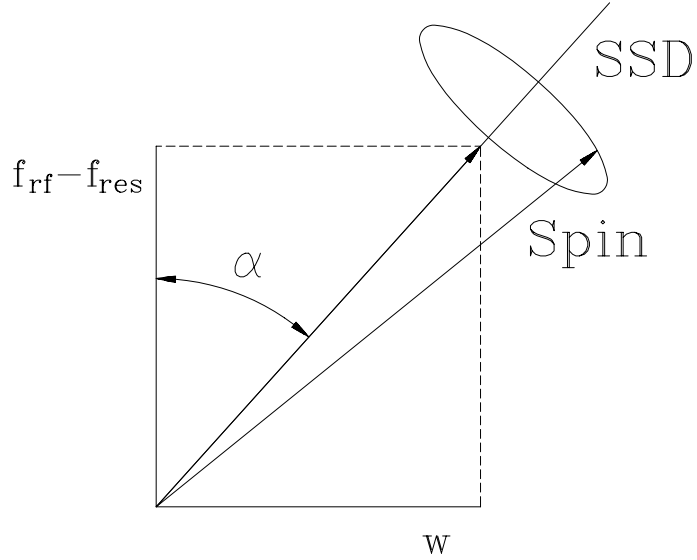}
\caption{\small Stable spin direction tilt in the rotating frame}
\end{figure}
tilted from vertical by the angle $\alpha$ given by:
\begin{equation}
cos(\alpha) = \frac{f_{rf} - f_{res}}{\sqrt{(f_{rf}-f_{res})^2 + (\epsilon\!f_c)^2}},
\end{equation}
where $f_{rf} = \frac{\omega_{rf}}{2\pi}$ is the rf magnet's frequency, 
$f_{res}$ is the resonance's central
frequency, $f_c$ is the circulation frequency, and $\epsilon$ is the resonance's strength, 
which is proportional to the integrated rf magnetic field's amplitude $\int{\!B_{rf}\!\cdot\!dl}$:
\begin{equation}
\epsilon = \frac{(1 + G)e}{\gamma m_pv}\frac{\int{\!B_{rf}\!\cdot\!dl}}{2}.
\end{equation}
Note that only one half of the
rf magnetic field amplitude contributes, because only one of the
two counter-rotating fields induces the resonance.

In the laboratory frame, this tilt would result in an apparent depolarization, since
the horizontal component of the polarization would average to zero.
If an rf magnet was turned on abruptly at frequency $f_{rf}$, the 
vertical beam polarization with an initial value $P_0$ would be reduced to:
\begin{equation}
P = P_0[cos(\alpha)]^2 = P_0\frac{(f_{rf} - f_{res})^2}{(f_{rf}-f_{res})^2 + (\epsilon\!f_c)^2},
\end{equation}
which is a first-order Lorentzian in the rf frequency.

However, it has been observed~\cite{width} that the actual 
measured width $w$ of an
rf depolarizing resonance is not always equal to its normalized strength $\epsilon\!f_c$, as
theory predicts; in fact, the measured width is frequently wider than the normalized
resonance strength. Moreover, the shape of the resonance curve does not
always follow a first-order Lorentzian. Several mechanisms for this resonance width 
enhancement have been proposed, including: a large beam energy spread, 
which could increase the resonance width for
either an rf solenoid or an rf dipole; and large coherent betatron oscillations, 
caused by an rf dipole. They will be discussed in Chapter~IV.

The frequency $f_{res}$ at which an rf depolarizing resonance is centered
can be predicted from Equation~I.7 by knowing the circulation frequency $f_c$ and
the spin tune $\nu_s$. Here we are mainly interested in the rf depolarizing resonances
in the presence of a full or a nearly full Siberian snake. 

First, consider a nearly full Siberian snake, whose snake strength $s$ is very
close but not exactly equal to 1. Let us assume that the 
snake strength is less than
1 by a very small amount $\delta s$. Then the spin tune, defined by Equation~II.53, is:
\begin{equation}
\nu_s = \frac{1}{\pi}cos^{-1}\left\{cos(\pi G\gamma)cos(\frac{\pi\!(1-\delta s)}{2})\right\}.
\end{equation}
For very small $\delta s$ this simplifies to:
\begin{equation}
\nu_s \simeq 1/2 - \frac{\delta s}{2}cos(\pi G\gamma). 	
\end{equation}
According to Equation~I.7, rf depolarizing resonances will occur at the frequencies:
\begin{equation}
f_{rf} = f_c\left\{n\pm[1/2-\frac{\delta s}{2}cos(\pi G\gamma)]\right\},
\end{equation}
where $n$ is an integer. Thus, for each $n$, there will be two resonances, 
which are closely spaced around 
$f_c(n+1/2)$ at frequencies:
\begin{equation}
f_l = f_c[n+(1/2-\frac{\delta s}{2}cos(\pi G\gamma))] = f_c[n+1/2-\frac{\delta s}{2}cos(\pi G\gamma)]
\end{equation}
and 
\begin{equation}
f_u = f_c[(n+1)-(1/2-\frac{\delta s}{2}cos(\pi G\gamma))] = f_c[n+1/2+\frac{\delta s}{2}cos(\pi G\gamma)];
\end{equation}
the subscript $l$ refers to a {\bf lower}-frequency resonance, and subscript $u$ 
refers to an {\bf upper}-frequency resonance. These resonances are separated by
\begin{equation}
\delta\!f = f_u - f_l = f_c\delta s~cos(\pi G\gamma).
\end{equation}
If this separation $\delta\!f$ is comparable to the resonances' widths, then 
the two resonances may overlap.

Now, consider the rf
depolarizing resonances frequencies in the presence of an exactly 100$\%$ Siberian
snake.
From Equation~II.48, the spin tune is then equal to exactly 1/2, and
expression for the lower and the upper resonance frequencies from Equations~II.66 and II.67
coincide:
\begin{equation}
f_l = f_u = f_c(n+1/2),
\end{equation}
where $n$ is an integer. Thus, with an exactly 100$\%$ snake, there would be a pair of
completely overlapping rf depolarizing resonances, located at each half-integer value of the
circulation frequency.

\section{Froissart-Stora Formula and Spin-Flipping}

Now consider a proton beam, stored in an accelerator ring with
an rf magnet, whose frequency $f_{rf}$ is very far from the resonance
frequency $f_{res}$; for example, let $f_{rf}=f_{res}-\Delta\!f/2$. 
The tilt angle $\alpha$ is then very small
and the polarization is nearly perfectly vertical. Now let the
rf magnet's frequency increase monotonously by some amount $\Delta\!f$
in some time $\Delta\!~t$, 
so that at some instant $f_{rf}=f_{res}$, and then later
$f_{rf}=f_{res}+\Delta\!f/2$. According to Equation~II.60,
the SSD, will then tilt from the vertical and will cross the horizontal plane when
$f_{rf}~=~f_{res}$; it will reverse to the opposite direction at
the end of the frequency ramp at time $\Delta\!~t$. If the frequency ramp
rate $\Delta\!f/\Delta\!~t$ is sufficiently slow, then the spin would follow the SSD, and flip.

Froissart and Stora~\cite{stora} derived an equation to 
relate the final beam polarization, $P_f$,  to the
initial polarization $P_i$, which was given earlier as Equation~I.5:
\begin{equation}
P_f = P_i[2 e^{- \frac{\mbox{\scriptsize $(\pi \epsilon f_c)^2$}}{\mbox{\scriptsize $(\Delta f / \Delta t)$}}} - 1];
\end{equation}
note that $\epsilon$ is the resonance strength and $\Delta\!f/ \Delta\!~t$ is the
frequency variation rate, where $\Delta\!f$ is the frequency range during the
frequency ramp time $\Delta\!~t$ (assuming a constant rate). As discussed
the Chapter~I, depending on the values of $\Delta\!f/ \Delta\!~t$ and the resonance
strength, such a ramp could cause either partial or full depolarization or
partial or full spin-flip. The final polarization $P_f$ is plotted against $\Delta\!~t$, assuming
$P_i = 1$, in Figure~II.3.
\begin{figure} [b!]
\epsfig{file=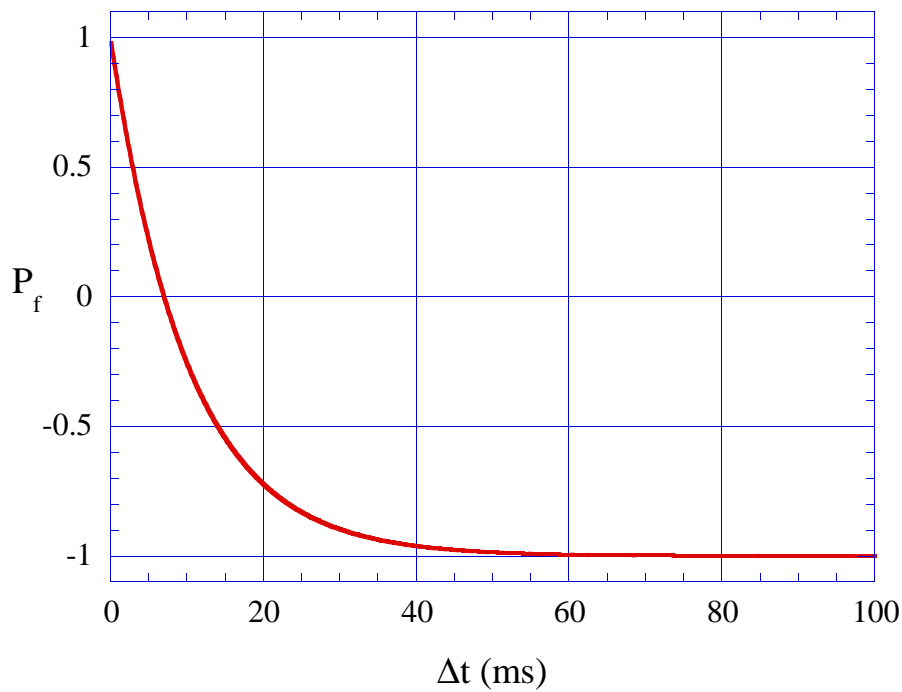}
\caption{\small Beam polarization after crossing an rf depolarizing resonance vs. ramp time}
\end{figure}

However, the Froissart-Stora formula is only exactly valid
for an isolated resonance, and for a frequency range varying from $-\infty$ to
$+\infty$. In practice, depolarizing resonances cannot be totally isolated, and the
frequency range cannot be infinite. Thus, the actual shape of a spin-flipping
curve may not agree with the Froissart-Stora formula prediction, which is shown in Figure~II.3. 
To maximize the spin-flip efficiency
\begin{equation}
\eta = \frac{-P_f}{P_i},
\end{equation}
the parameters $\epsilon$, $\Delta\!f$, and $\Delta\!~t$ must be carefully adjusted.

\chapter{ Experimental Apparatus}

\section{Introduction}

Our spin-flipping experiments were done at the Indiana University Cyclotron 
Facility Cooler Ring during 1997-1998. Much of the apparatus used in these experiments
was described in earlier 
papers~\cite{tune,width,adk,goodwin,minty,anferov,baiod,phelps,blinov,caussyn,ohmori,alexeeva,crandell,
phelps2,blinov2,chu,blinov3,goodthe,mintthe,widththeory,chuthe}. The layout of IUCF during that period
is shown in Figure~III.1. 

The polarized
protons from the ion source were first accelerated in the Injector Cyclotron
and next in the Main Cyclotron. The spin precession "$\theta$''-solenoid 
rotated the injected beam polarization direction to match 
it with the 
SSD in the Cooler Ring. We monitored this injected beam polarization using 
the Beam Line 3 (BL3) polarimeter. Then, the beam was injected and stacked
in the Cooler Ring for 10 to 30 seconds to achieve a high beam current.

The stored beam's polarization was measured using a carbon skimmer target
polarimeter in the Cooler Ring's A region. 
The polarization direction of the injected beam was reversed
every injection cycle to reduce the systematic error in this polarization
measurement.

\section{Beam Acceleration in Cyclotrons and Polarization Rotation}

The polarized proton beam was produced by the High Intensity Polarized
Ion Source (HIPIOS)~\cite{hipios}. Its 20~keV polarized protons were accelerated in the Cockroft-Walton 
pre-accelerator to 600~keV and transported to the Injector Cyclotron, where they were
\begin{figure}
\epsfig{file=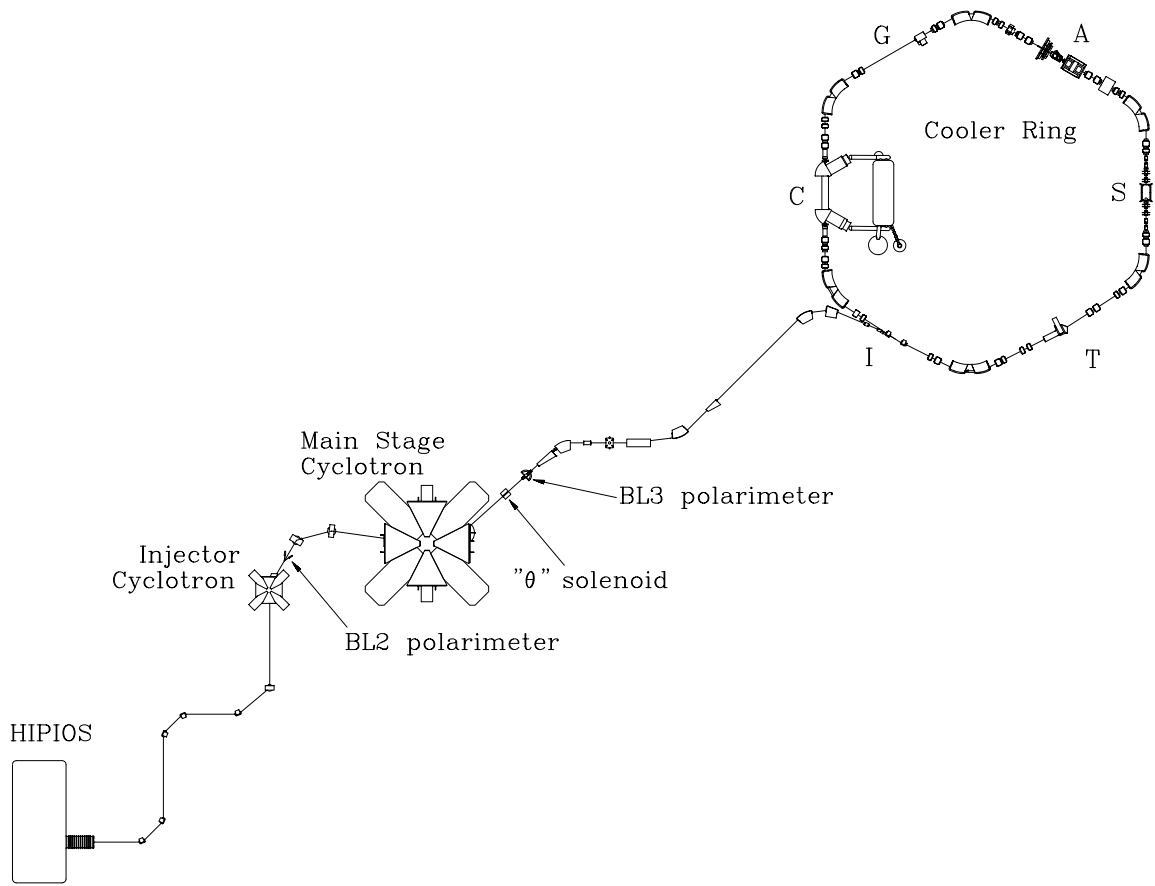}
\caption{\small Indiana University Cyclotron Facility circa 1997-1998.}
\end{figure}
accelerated to 15~MeV and then transported to the Main Stage Cyclotron,
which accelerated the beam to 104.1~MeV.

The beam polarization emerging from the polarized ion source was vertical to
match the field direction in the cyclotrons. 
However, with a full Siberian snake present in the Cooler Ring, the SSD
was horizontal everywhere in the Ring. To match the injected polarization
direction with this horizontal SSD,
the vertical polarization was rotated through 90$^{\circ}$ into the horizontal plane
in the ``$\theta$'' solenoid magnet, which required about 33.1~Amps 
for this 90$^{\circ}$ rotation at 104.1~MeV.

\section{Polarized Beam Injection into the Cooler Ring}

The polarized beam was stack-injected using: a Lambertson septum magnet in the middle
of the Cooler Ring's I section; a pair of fast-rise kickers at either end of this section;
and a pair of RF cavities, as shown in Figure~III.2.
\begin{figure} [b!]
\begin{center}
\epsfig{file=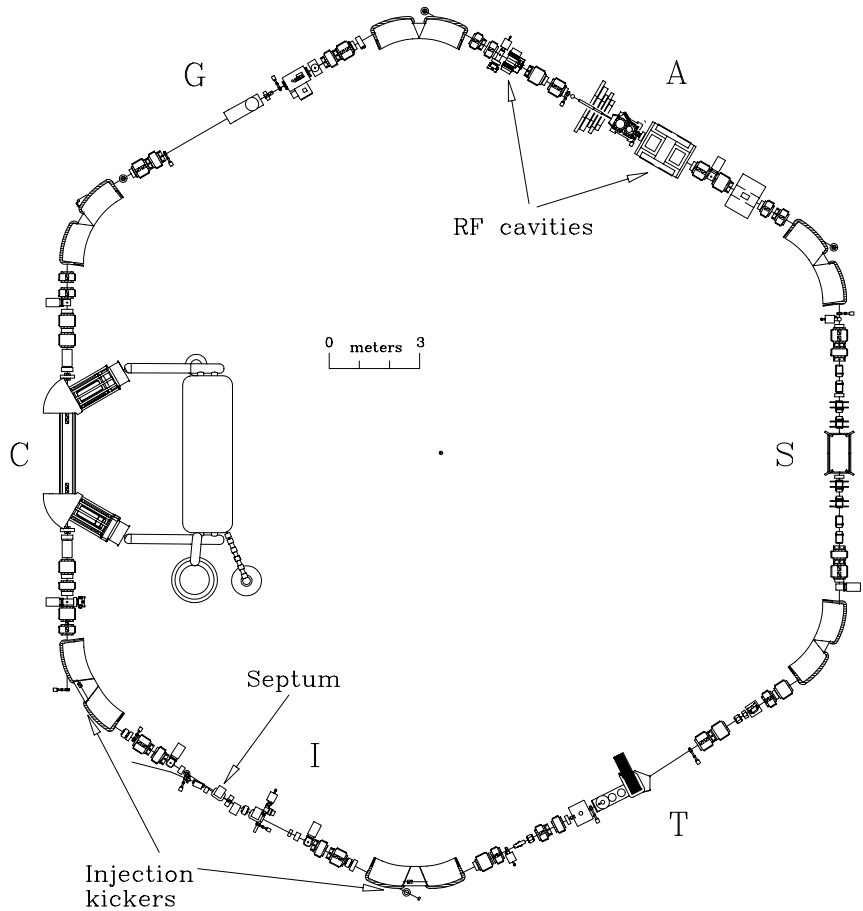}
\end{center}
\caption{\small Injection hardware in the Cooler Ring}
\end{figure}

When the two kickers were on, the injected beam entered the ring and 
circulated
around the {\bf injection orbit}, which passes through the septum 
and is slightly different from the {\bf stored beam orbit} 
as shown in Figure~III.3. Abruptly turning off the kickers then moved the 
injected beam into the {\bf stack orbit} in the field-free region of the Lambertson magnet.
\begin{figure} [h!]
\begin{center}
\epsfig{file=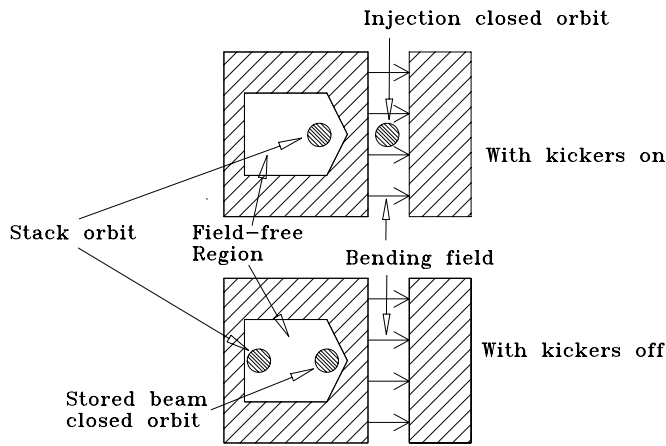}
\end{center}
\vspace{-0.3 in}
\caption{\small Beam positions in the Lambertson septum}
\end{figure}
\vspace{-0.15 in}

The injected protons had a slightly higher energy than the stored protons; thus the
{\bf stack orbit's} radius was slightly larger than the {\bf stored beam orbit}.
The rf cavities were then used to quickly decelerate the newly injected 
protons and thus move them to the {\bf stored beam orbit}.

The Cooler Ring's electron cooling system was then used to reduce the protons' 
momentum spread by injecting into the Cooling straight section C a nearly 
mono-energetic beam of electrons, whose velocity matched the stored
protons' velocity. Because of their much lower mass, the electrons had
a much lower energy spread and thus were "cooler" 
than the more massive protons. The "hot" protons then interacted 
with the "cold" electrons,
which reduced the protons' momentum spread~\cite{cooling}. 
The "heated" electrons were then removed from the Cooler Ring.

The protons were cooled for many turns with the injection kickers off
until they were cooled such that their momentum spread was about 
$\Delta p/p=2\cdot10^{-4}$. 
This injection and cooling process was repeated about 
every 150~$ms$, until the desired beam current was stored.

This stacking injection resulted in bunched beam 
currents of up to 500~$\mu A$ in the Cooler Ring by using the DC proton 
beam from the cyclotrons. 
However, our full Siberian snake caused
some beam aperture difficulties (see Section~III.5), and 
limited the stored beam current in the Cooler Ring to about 10~$\mu A$; fortunately this was
adequate for a $\pm~2\!$~-~$\!3~\%$ measurement of the polarization in about 15 minutes.

\section{The Cooler Ring}

The full layout of the Cooler Ring is shown in Figure~III.4. The Ring's circumference
is about 82.78~m and its maximum accelerated proton beam energy is 
\begin{figure} [h!]
\vspace{-.15 in}
\epsfig{file=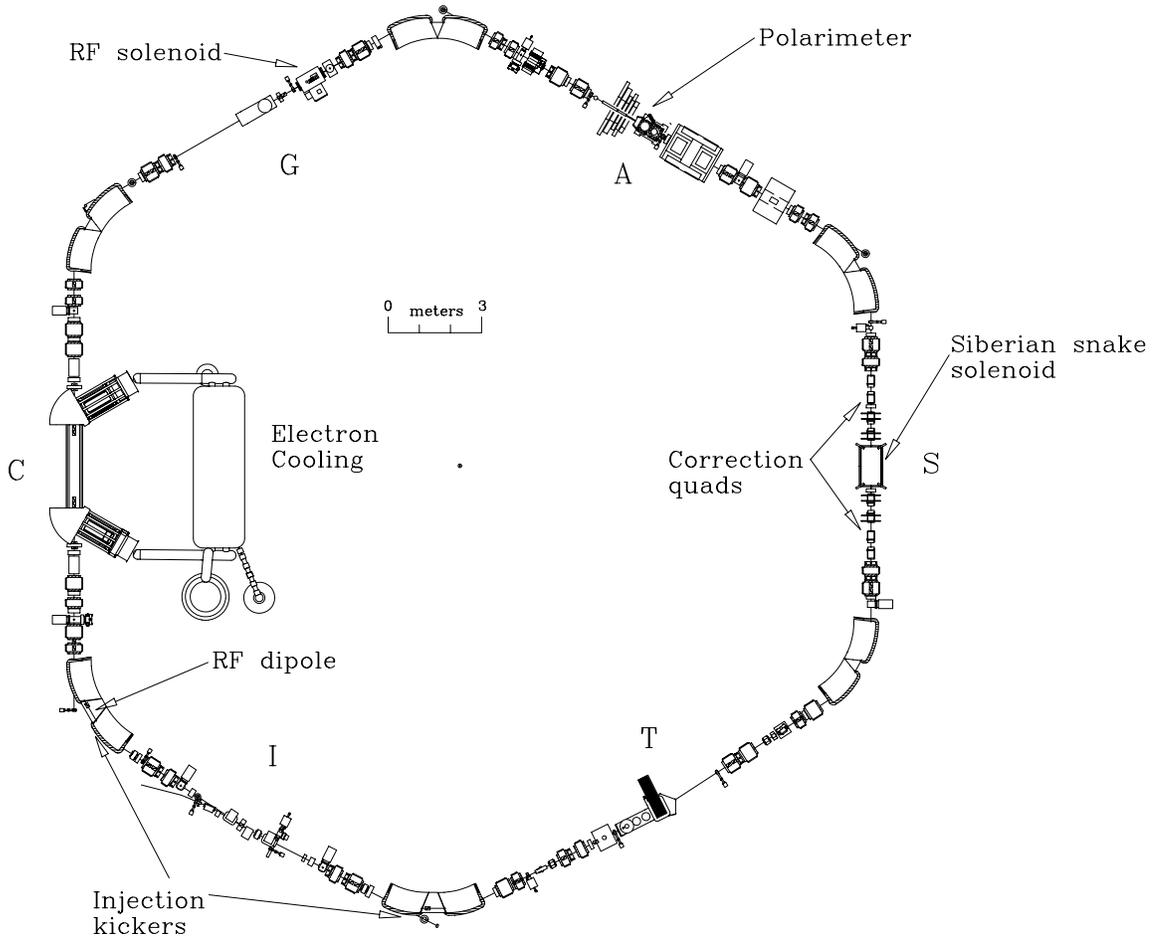}
\caption{\small The Cooler Ring}
\end{figure}
about 500~MeV; the Ring's maximum injection energy is determined by the
Main Stage Cyclotron's maximum energy, which is about 200~MeV.  
The Cooler Ring
is a strong-focusing synchrotron with a 3-fold symmetry and six $\pi/3$
bends between its six straight sections; it has a FOFDOD quadrupole lattice. 
The Ring's main quadrupoles are organized into several "combos" with separate
power supplies; this allows the Cooler Ring's betatron tunes to be varied. 
The accelerating rf field is provided
by the larger of the two rf acceleration cavities, which were 
mentioned in Section~III.3. This cavity, which is called the PPA cavity,
generated the rf acceleration field and provided beam bunching by
running at an appropriate harmonic of the beam's circulation frequency $f_c$.
The smaller rf cavity, called the MPI cavity, was only used during the injection;
its frequency was equal to $f_c$.

The Cooler Ring's six
straight sections accommodate a large amount of experimental hardware
for both nuclear physics and accelerator physics experiments.
The {\bf injection section} I was described in detail in Section~III.3.; its
upstream injection kicker magnet was also used as an rf dipole for our experiment. 
The T section is dedicated to nuclear physics experiments, while the Siberian snake solenoid 
and its eight correction quadrupoles are installed in the {\bf snake section} S. 
The polarimeter is located in the A section, along with some nuclear physics
experiments.
The G section accommodates the RF solenoid and a small polarized hydrogen target,
used in three-body forces studies. The Electron Cooling System 
is installed in the {\bf cooler section} C as dicussed in Section~III.3.

\section{The Siberian Snake}

A Siberian snake can be constructed of either a single solenoid magnet or a 
series of dipole magnets with different field orientations; however, a solenoid 
is most suitable for low energies. Although its $\int\!B\!\cdot\!dl$
is energy dependent, it does not create any orbit excursions, which can be
very large inside a dipole snake at low energies. Moreover, for energies below 
500~MeV, a solenoidal snake containing a superconducting solenoid
can be much shorter than a dipole snake. Thus, at the 500~MeV IUCF Cooler Ring, 
a superconducting solenoid magnet was used as our snake.
\begin{figure}[t!]
\epsfig{file=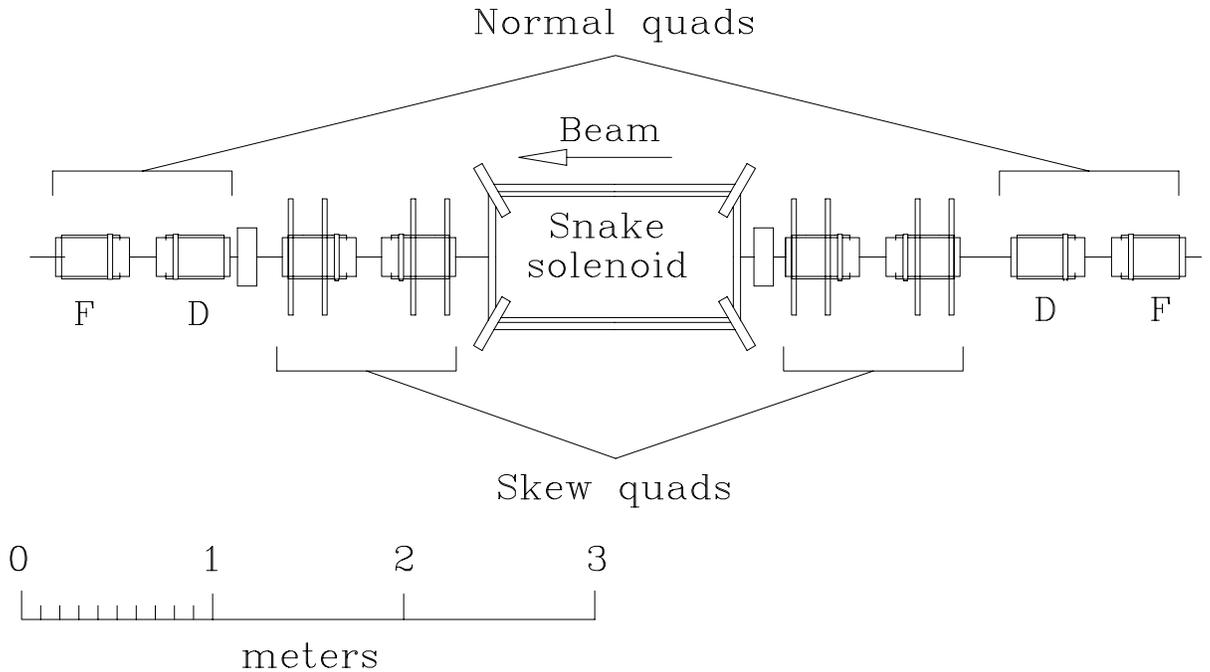}
\vspace{-0.5 in}
\caption{\small The Siberian snake}
\end{figure}

One unfortunate feature of a high-field solenoid is that it causes strong beam focusing
and strong betatron coupling. To allow successful beam operation, these two effects 
were compensated using a set of 8 correction
quadrupoles. The Siberian snake's solenoid and the quadrupoles are shown in
Figure~III.5. 

The eight correction quadrupoles 
consisted of two antisymmetric focusing-defocusing pairs of normal quadrupoles far from
the solenoid and two antisymmetric pairs of skew quadrupoles near the solenoid.
Note that a skew quadrupole is a normal quadrupole, which is rotated by
some angle about its axis.
The normal quadrupoles' strengths were set to cancel the solenoid's focusing
of the beam, while the skew quadrupoles' angles and currents were set to
cancel the betatron coupling caused by the solenoid. However, 
we could not completely cancel
these two effects; this caused a lower
beam intensity in the presence of the full snake.
\begin{figure}[b!]
\epsfig{file=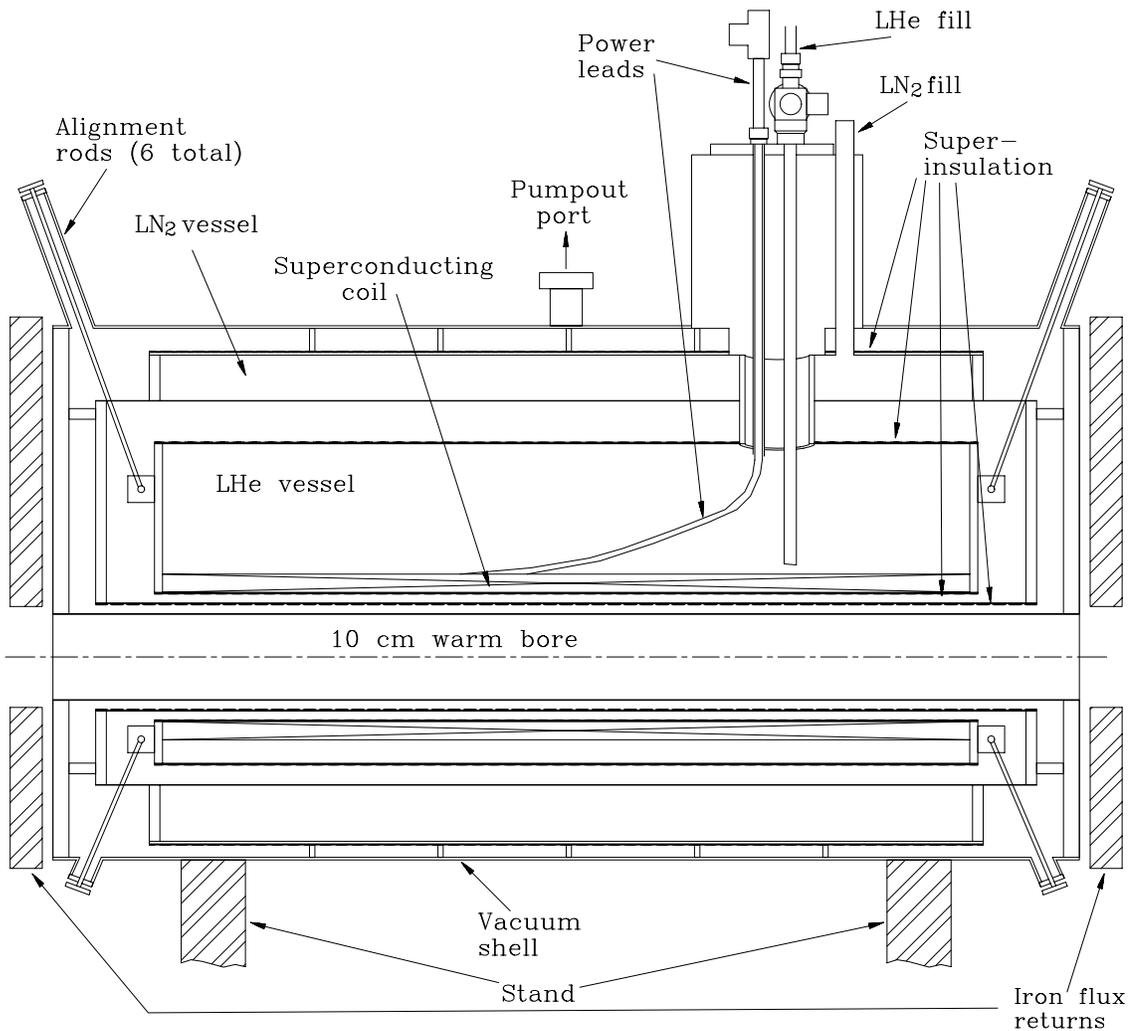}
\caption{\small The Siberian snake solenoid layout}
\end{figure}

The superconducting solenoid magnet had $N$~=~13,450 turns 
and could operate at a maximum current $I$~=~171~Amps, where its 
maximum field integral is:
\begin{equation}
\int\!B\!\cdot\!dl = \mu_0 NI = 2.90~ T\!\cdot\!m,
\end{equation}
where $\mu_0$ = 4$\pi\!10^{-7}~T\!\cdot\!m\!\cdot\!A^{-1}$.
For a 100~\% Siberian snake ($s = 1$) at 104.1~MeV, 
where the proton's momentum is 0.454~GeV/c, Equation~II.38 indicates
that the required field integral is:
\begin{equation}
\int\!B\!\cdot\!dl = 3.752\cdot 1\cdot 0.454 = 1.70~T \cdot m,
\end{equation}

A schematic drawing of the Siberian snake's superconducting solenoid, 
used during 1997-1998, is shown in Figure~III.6.
The solenoid consists of a 122~cm long, 64~cm O.D. stainless steel vacuum shell
with a 10~cm warm bore,
a 50~liter liquid nitrogen tank, and an 80 liter liquid helium tank which contains
the solenoid coil. The liquid helium vessel is supported by six alignment
rods. All outer surfaces of the liquid nitrogen and liquid helium tanks are covered
with about 10 layers of superinsulation. The instrumentation rack on the top of the
vacuum shell houses the liquid helium and liquid nitrogen fill ports, and the current leads.
These power leads are cooled by the flow of very cold helium gas, evaporating from the
liquid helium tank. The solenoid is enclosed in a flux return box made of a 4~cm thick iron.

To ensure good stability of the snake strength and thus the 
spin tune during the run, we used a very stable 
EMI Corp. EMS 20-250 DC power supply for our snake solenoid; the supply's current
stability was about $\pm$~0.01$\%$. The power supply
could be controlled both locally and remotely; we used the remote mode
to adjust the current during the run.

The superconducting solenoid could operate for approximately 16 hours between 
refills of liquid helium. Since the snake solenoid could not operate 
in the persistent mode, each refill required ramping down the solenoid
current. After filling the solenoid with helium, it was necessary
to ramp it back up and then attempt to reset the current to its original value.
The snake solenoid's current was measured and set using a high precision DC current
transformer (DCCT). The output voltage of the DCCT, which was directly
proportional to the snake's current, was precisely
measured by a Keithley~155 nulling voltmeter; the precision of this
current measurement was about 10~p.p.m. 

\section{The RF-Solenoid}
\begin{figure}[t!]
\epsfig{file=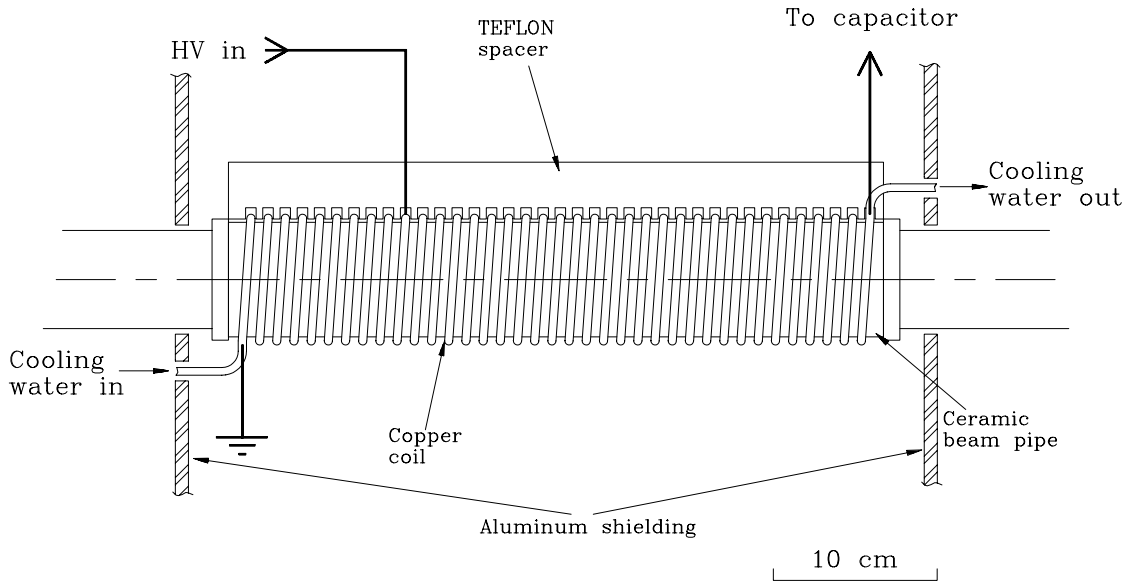}
\caption{\small RF solenoid}
\end {figure}
An rf resonance can be created by a magnet, whose field amplitude oscillates
at an "rf " frequency in the megahertz range. Such an rf magnet could be either an rf dipole
or an rf solenoid. 
Here, as with a Siberian snake, a solenoid
has several advantages at low energies. Although its strength is reduced by
the Lorentz contraction of its longitudinal $\int\!B\!\cdot\!dl$, this longitudinal field does not
affect the beam motion. An rf dipole, at a frequency near any integer multiple 
of the horizontal or vertical betatron frequencies, could significantly increase the
transverse beam oscillations, which could destroy the beam.
\begin{figure} [b!]
\epsfig{file=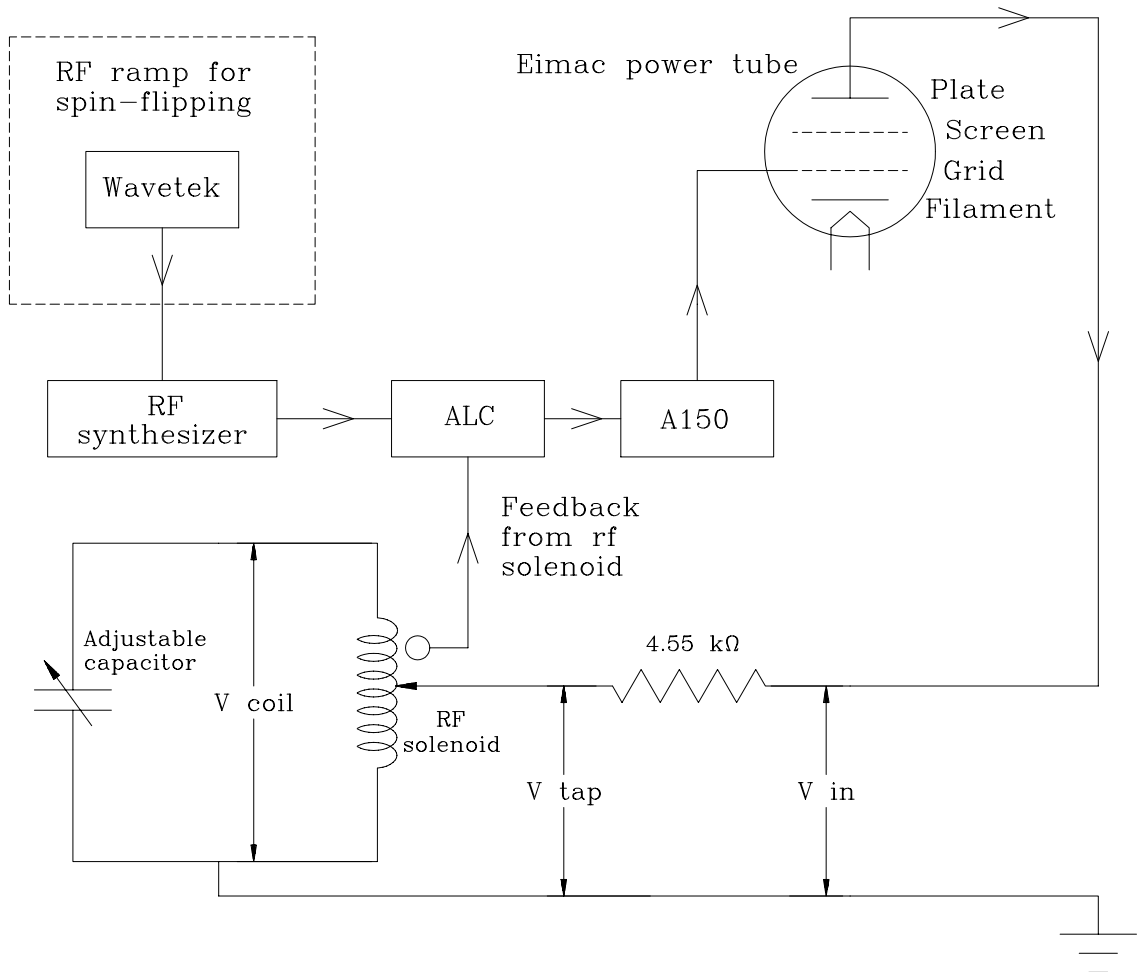}
\caption{\small RF solenoid's electronics}
\end {figure}

The rf solenoid magnet, which was installed in the Cooler Ring, 
is shown
in Figure~III.7~\cite{widththeory}. It consists of a 37-turn water-cooled copper coil, wound 
around a 3'' O.D. ceramic beam pipe, 
with 3 Teflon combs to separate the adjacent turns of the coil, and
an aluminum housing, which
shields the surrounding equipment and nearby radios from the solenoid's rather strong rf radiation.

A diagram of the rf solenoid's electronic components is shown in Figure~III.8.
The rf signal for the rf solenoid was initially generated using a Hewlett-Packard
HP8656 frequency synthesizer; a typical signal's frequency was about 2.2~MHz and its
amplitude was typically set to +6~dBm, which corresponded to about 0.9~V peak-to-peak
voltage with the rf synthesizer's 50~Ohms load. 
This signal's amplitude was then stabilized by an automatic level controller box (ALC), 
which made the rf solenoid's voltage independent of frequency by measuring the rf field with
a pick-up loop inside the rf solenoid and then adjusting the amplification to keep the field constant. 
The signal was then amplified by an ENI Inc. A150 rf power amplifier. 
The output of the A150 was applied to the
grid of the EIMAC 4CW 10,000A tetrode amplifier tube, which produced an rf
voltage of up to about 7~kilovolts.

We ran the rf synthesizer in a fixed frequency mode for our rf resonances studies.
For the spin-flipping studies, the rf synthesizer's frequency was modulated to
produce a frequency ramp around the 2.2~MHz central frequency.
A ramp's typical range $\Delta\!f$ was about $\pm$6~kHz, its time duration
$\Delta\!~t$ was between 1 and 1000~ms. 
This modulation was produced by a Wavetek model 90 arbitrary waveform generator.

To achieve a high rf magnetic field with this 37-turn solenoid, a 
10-1000~pF vacuum variable
capacitor was wired in parallel with the solenoid coil to form an LC resonant circuit.
The rf voltage $V_{in}$ from the EIMAC power amplifier tube
was applied to the coil's tap point through a 4.55~$k\Omega$ resistor to increase
the quality factor of the LC resonance circuit; the typical Q-value at 2.2~MHz was about 200. 
The tap point was located at approximately
1/4 of the length of the coil to match the input impedance of the resonance circuit to the
output impedance of the tube. Thus, the total rf voltage amplitude across 
the solenoid $V_{coil}$ was about 4 times higher than the applied voltage $V_{tap}$. We ran
at the maximum rf voltage across the solenoid of about 6~kV amplitude; its maximum longitudinal 
$\int\!B\!\cdot\!dl$ was about 1.6~T$\cdot$mm.

\section{The RF-Dipole}

To study rf depolarizing resonances with an rf dipole, we used one of the
Cooler Ring's injection kicker dipoles; 
\begin{figure} [t!]
\begin{center}
\vspace{-0.2 in}
\epsfig{file=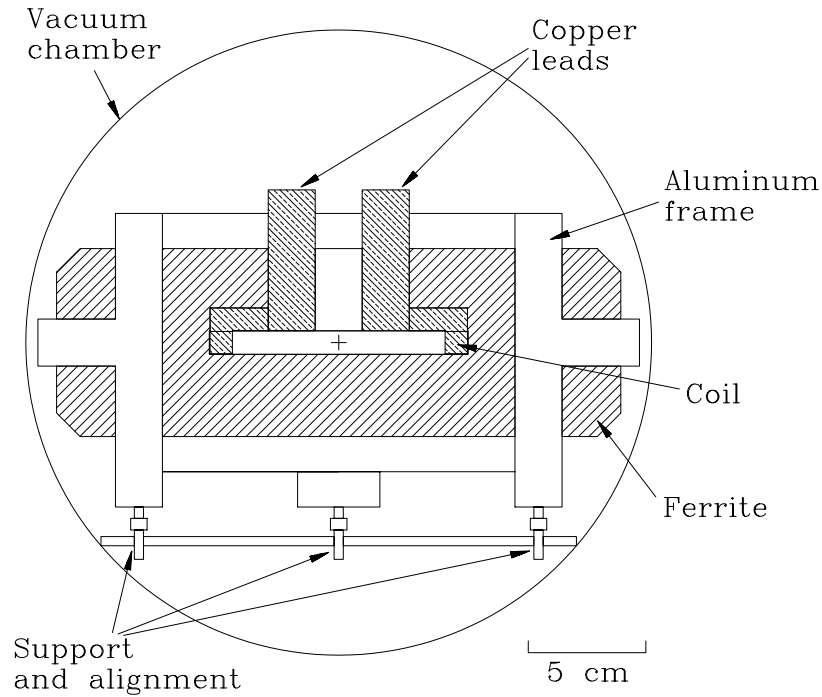}
\end{center}
\vspace{-0.4 in}
\caption{\small RF dipole (beam view)}
\end{figure}
\begin{figure}[b!]
\vspace{-0.2 in}
\epsfig{file=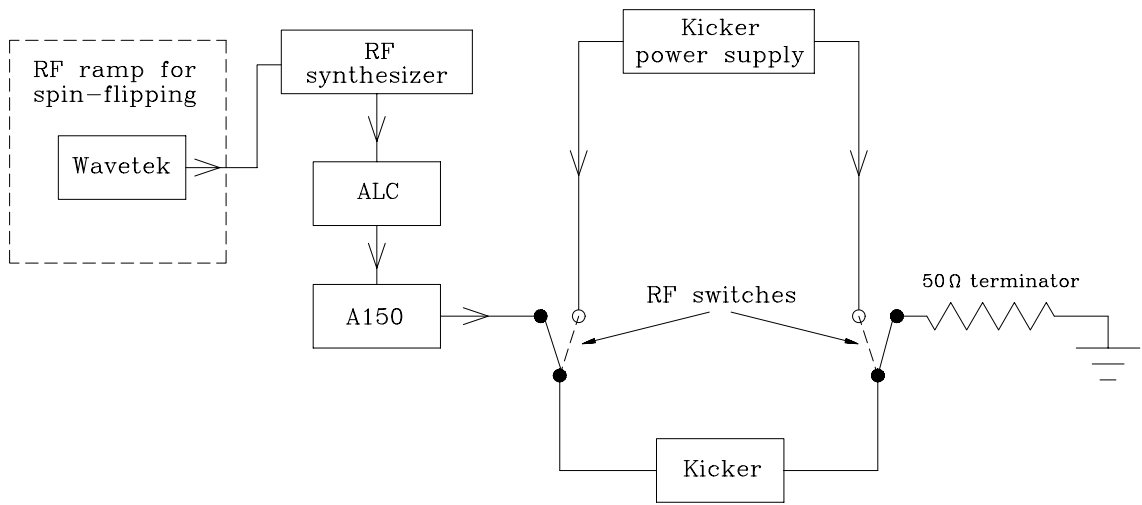}
\caption{\small RF dipole's electronics}
\end{figure}
the kicker magnet is shown in Figure~III.9. It is a ferrite magnet
8~cm high by 24~cm wide by 40~cm long 
with a one-turn coil; it fits inside a 26~cm O.D. stainless steel vacuum chamber
with a 1.6~mm wall thickness.

A diagram of the rf dipole's electronics is shown in Figure~III.10.
Since, during each Cooler Ring cycle, the ferrite dipole had to be used as both
an injection kicker and an rf dipole, it was necessary
to switch the dipole's input between the kicker power supply and
the rf power supply. Therefore, we installed two pneumatic rf switches
with a 100~msec switching time.
During injection, the switches were in the inner position, and the
kicker power supply pulses were applied to the dipole. At the start of the flattop,
the switches were turned to the outer position, and the output of 
an A150 rf power 
amplifier was connected to one lead of the dipole.
The other lead was terminated with a 50~Ohm resistor to
match the output impedance of the A150. The rf signal for the
rf dipole was generated by the same rf synthesizer and ALC
as was used for the rf solenoid. However, we  did not use the EIMAC
power amplifier, since its high-voltage output might have destroyed
the L~=~4.1~$\mu$H ferrite dipole. Moreover, the rf dipole
was not a part of a high-Q resonant LC circuit.

Unfortunately, we were not able to perfectly match the output of the A150
power amplifier and the input of the rf dipole. This mismatch, which was probaby caused by
the rf switches, significantly reduced 
the voltage across the dipole and consequently its resonance strength.
The maximum r.m.s. voltage that we were able to reach was only about 23~V;
this corresponded to an $\int\!B\!\cdot\!dl$ of about 0.03~T$\cdot$mm.

\section{Polarimetry}

The proton beam polarization was monitored carefully during each stage of the beam
acceleration. The initial polarization of the beam emerging from the Injector Cyclotron
was measured by the BL2 polarimeter, which used the elastic scattering of protons
on its $^4$He gas target~\cite{bl2}. The analyzing power of $p~+~^4He\rightarrow~p~+~^4He$ 
at the Beam-Line-2 energy of about 15~MeV was close to 1.0 for all angles.

The polarization of the beam extracted after acceleration by the Main
Cyclotron was measured by the BL3 polarimeter~\cite{bl3}, which is shown in Figure ~III.1. 
The BL3 polarimeter used a thin 0.5~$\mu\!\!~g~cm^{-2}$ deuterated
polyethylene film, and detected both the scattered protons and recoil 
deuterons at the angles of $73.6^{\circ}$ and $37.8^{\circ}$, respectively, in 
the up-down and left-right directions. The analyzing power $A_n$ varies 
with the incident protons' energy, with a typical value of about 0.6. 

The Cooler beam polarization was measured by a carbon target polarimeter in the A
region, which is shown in Figure~III.11~\cite{coolpol}. It consisted of a 3~mm thick 
carbon skimmer target, a collimator and
the detectors: two sets of wire chambers, u-v and x-y, and two sets of 
scintillators, F and E. The
skimmer target was moved into the circulating beam with its feedback
loop connected to the detector rate meter; an increased rate would slow
the inward motion of the target, thus providing a uniform rate
of data acquisition during the polarization measurements.
\begin{figure}[t!]
\vspace{.25 in}
\epsfig{file=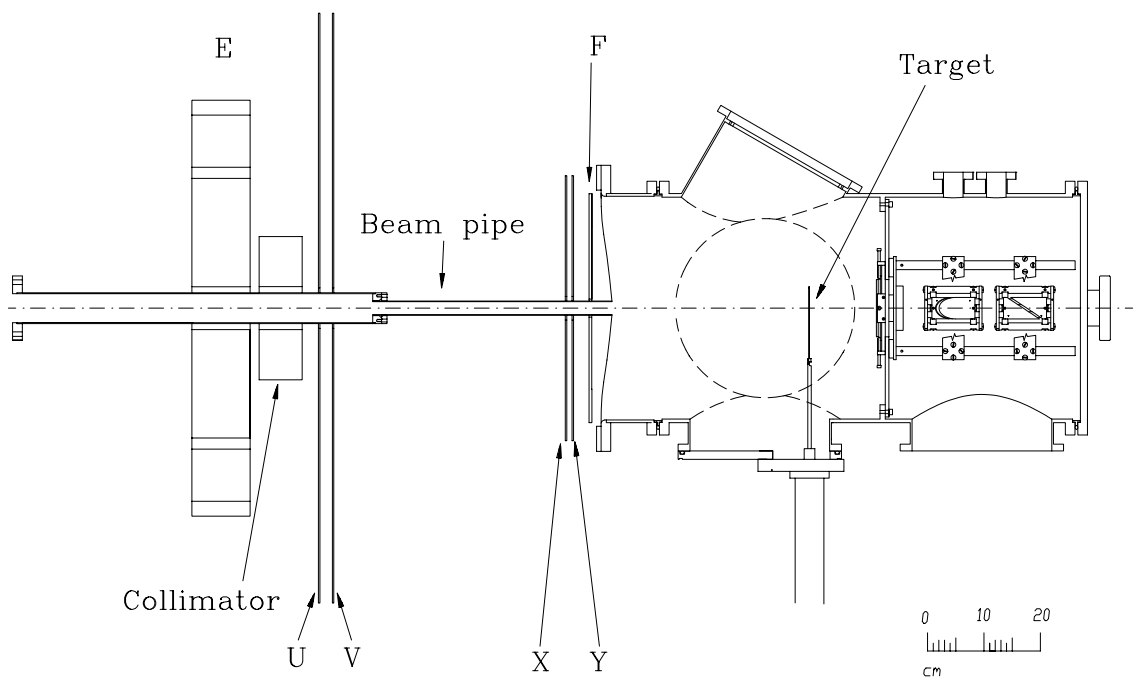}
\caption{\small Cooler polarimeter (top view).}
\end{figure}

The polarization measurement was based on the asymmetry
of 104.1~MeV protons elastically scattered on $C^{12}$ with an 
average analyzing power of about
0.233. The trigger for an elastic event was provided by an F and E coincidence,
while the orthogonal wire chambers determined the tracks of scattered particles,
ensuring that only events originating at the target were counted.
These wire chambers also measured the angle $\theta$ between the scattered
proton's track and the beam axis. 

The elastic events were then divided, using software on-line analysis, into four regions
(up, down, left and right) of $\pi$/2 each, and the vertical polarization was calculated by
\begin{equation}
P = \frac{1}{\bar{A}_n}\frac{\sqrt{N_{L\uparrow}N_{R\downarrow}}-\sqrt{N_{L\downarrow}N_{R\uparrow}}}{\sqrt{N_{L\uparrow}N_{R\downarrow}}+\sqrt{N_{L\downarrow}N_{R\uparrow}}},
\end{equation}
where $\bar{A}_n$ is the angle weighted averaged analyzing power,
and $N_{L(R)\uparrow(\downarrow)}$ are the total numbers of events scattered into the $Left (Right)$
detectors with the injected beam polarization $up (down)$.
Similarly, the radial beam polarization was calculated from:
\begin{equation}
P = \frac{1}{\bar{A}} \frac{\sqrt{N_{U\uparrow}N_{D\downarrow}}-\sqrt{N_{U\downarrow}N_{D\uparrow}}}{\sqrt{N_{U\uparrow}N_{D\downarrow}}+\sqrt{N_{U\downarrow}N_{D\uparrow}}},
\end{equation}
where $N_{U(D)\uparrow(\downarrow)}$ are the total numbers of events scattered 
into the $Up (Down)$ detectors with the injected beam polarization $up (down)$.

This algorithm of measuring the polarization eliminates most 
systematic errors, such as unequal efficiencies of different detectors 
and beam current fluctuations from cycle to cycle, which otherwise could
cause false asymmetries. The main source of the measured polarization's 
systematic error is then the value of the analyzing power. 

The analyzing power for proton-carbon elastic scattering was measured by several 
groups~\cite{ay1,ay2,ay3,McN}
at many energies and angles. We determined the analyzing power using the parameterization of McNaughton 
{\it et al.}~\cite{McN} for each momentum $p_{lab}$ and laboratory scattering angle
$\theta_{lab}$:
\begin{equation}
A_n(\theta_{lab}, p_{lab}) = \frac{ar}{1 + bp_{\perp}^2 + cp_{\perp}^4},
\end{equation}
where $p_{\perp} = p_{lab}\sin{\theta_{lab}}$, the momentum $p_{lab}$ is in $GeV/c$, and the 
parameters $a$, $b$ and $c$ are given by
\begin{equation}
a  = a_0 + a_1p' + a_2p'^2 + a_3p'^3 + a_4p'^4,
\end{equation}
\begin{equation}
b  = b_0 + b_1p' + b_2p'^2 + b_3p'^3 + b_4p'^4,
\end{equation}
\begin{equation}
c  = c_0 + c_1p' + c_2p'^2 + c_3p'^3 + c_4p'^4,
\end{equation}
where $p' = p_{lab} - 0.7~GeV/c$, and the parameters $a_i$, $b_i$ and $c_i$ 
are given in Table III.1.
\begin{table}[h!]
\begin{center}
\begin{tabular} {|c|ccccc|}
\hline
& 0 & 1 & 2 & 3 & 4 \\
\hline
$a_i$ & 5.3346 & -5.5361 & 2.8353 & 61.915 & -145.54 \\
$b_i$ & -12.774 & -68.339 & 1333.5 & -3713.5 & 3738.3 \\
$c_i$ & 1095.3 & 949.50 & -28012 & 96833 & -118830 \\
\hline
\end{tabular}
\caption{\small McNaughton {\it et al.}\cite{McN} parameterization coefficients for $A_n$,
which are valid 
between 100 and 450~MeV}
\end{center}
\end{table}

\chapter{ Experimental Results and Analysis}

\section{Introduction}
The results of four experimental runs of CE-69~\cite{ce69}
are discussed here. These include measurements of the locations and widths of 
snake rf depolarizing resonances using both the rf solenoid and the rf dipole; 
they also include spin-flipping studies using these rf magnets in the presence of 
a nearly-full Siberian snake.

For all four runs, a 104.1~MeV polarized proton beam was stored in the IUCF
Cooler Ring;
the circulation frequency $f_c$ was 1,504,900$\pm$10~Hz. The Ring's rf acceleration
cavity operated at the sixth harmonic of $f_c$ at a frequency of 9,029,400$\pm$10~Hz.

The Siberian snake solenoid's current varied slightly from run to run around 103~A;
this corresponded to a snake strength of approximately 1.02.

\section{RF resonances}
In order to spin-flip a stored polarized beam using an rf depolarizing resonance, 
the resonance's location and width must be first determined;
then the rf frequency ramp's central frequency and
frequency range $\Delta\!f$ can be chosen. With a nearly-full Siberian snake in the Cooler Ring, 
the spin tune was very close to 1/2, as given by 
Equation~II.63. Recall that Equations~II.66~and~II.67 indicate that rf depolarizing resonances
should then be found in closely spaced pairs around half-integer values of the circulation
frequency $f_c$. In our experiments we mainly studied the rf 
depolarizing resonances near $f\!~=\!~f_c(2-\nu_s)$. At 104.1~MeV, 
where $G\gamma=1.9918\simeq~2$, Equations~II.66~and~II.67 simplify to:
\begin{equation}
f_l\simeq f_c(1.5-\delta\!\!~s/2)
\end{equation}
and 
\begin{equation}
f_u\simeq f_c(1.5+\delta\!\!~s/2),
\end{equation}
where $\delta\!\!~s\!~=\!~|1-s|$ is the amount by which the snake strength $s$ differs from a full
snake.

The exact location of these "upper" and "lower" resonances depends strongly on the snake strength $s$,
which is proportional to the snake solenoid's current, as defined by Equation~III.39. 
As given by Equation~II.68, with a snake strength of about 1.02, the two resonances 
centered around 1.5$f_c$ would be separated by about $\delta\!f\simeq~30$~kHz.
This ensured that there was no overlapping between the upper and lower resonances.

\subsection{Experimental procedure}
For the rf resonance mapping, the horizontally polarized proton beam was injected into
the Cooler Ring, where it was stored and cooled. The rf solenoid or the rf dipole was turned 
on at a fixed frequency 
about 0.5~second after the start of the flattop. The data acquisition system 
was then turned on; then the target
was moved into the beam and the polarization measurements were started. 
The rf magnet remained on throughout the flattop.
At the end of the flattop, the data acquisition was turned off, then the rf magnet was turned
off, and the beam was kicked out of the Ring using one of the injection kickers. 
A typical timing pattern of the rf magnets, detectors and data acquisition for
an rf resonance mapping run is shown in Table~IV.1.
\begin{table}
\vspace{0.15 in}
\begin{center}
\begin{tabular}{|l|c|c|}
\hline
Timing for: & Time ON (sec) & Time OFF (sec) \\
\hline
Flattop & 0.000 & 30.000 \\
RF solenoid (dipole) & 0.600 & 29.950 \\
Wire chambers & 1.000 & 29.900 \\
Target & 1.600 & 29.900 \\
DAQ & 1.600 & 29.900 \\
\hline
\end{tabular}
\end{center}
\caption{\small Typical electronics timing for rf depolarizing resonance study
with a 30 second flattop.}
\end{table}

We first operated the rf solenoid
at its maximum strength to find the location of the rf depolarizing resonance.
We varied the rf solenoid's frequency in 1~kHz steps around the expected 
position of the resonance; if the resonance was observed, then we took
another set of data points with 1~kHz steps halfway between the first set of points; 
additional points were then taken near the resonance to better map it. Sometimes, 
the resonance was not found in the first set of points; we then scanned
the regions above and below the expected position with the 1~kHz steps until a
depolarization was observed.

This knowledge of the exact location of the rf resonance was used in subsequent rf dipole studies;
the rf dipole's resonance would otherwise be relatively hard to find, because it was a much 
weaker magnet.

\subsection{Data and Analysis}
Figure~IV.1 shows the rf solenoid depolarizing resonance data for each of the four experimental
runs. On each graph, the radial beam polarization is plotted against the rf solenoid's
frequency. Note the different central resonance frequency for each run; this was due
to slightly different snake solenoid currents.
\begin{figure}
\vspace{-0.2in}
\begin{center}
\epsfig{file=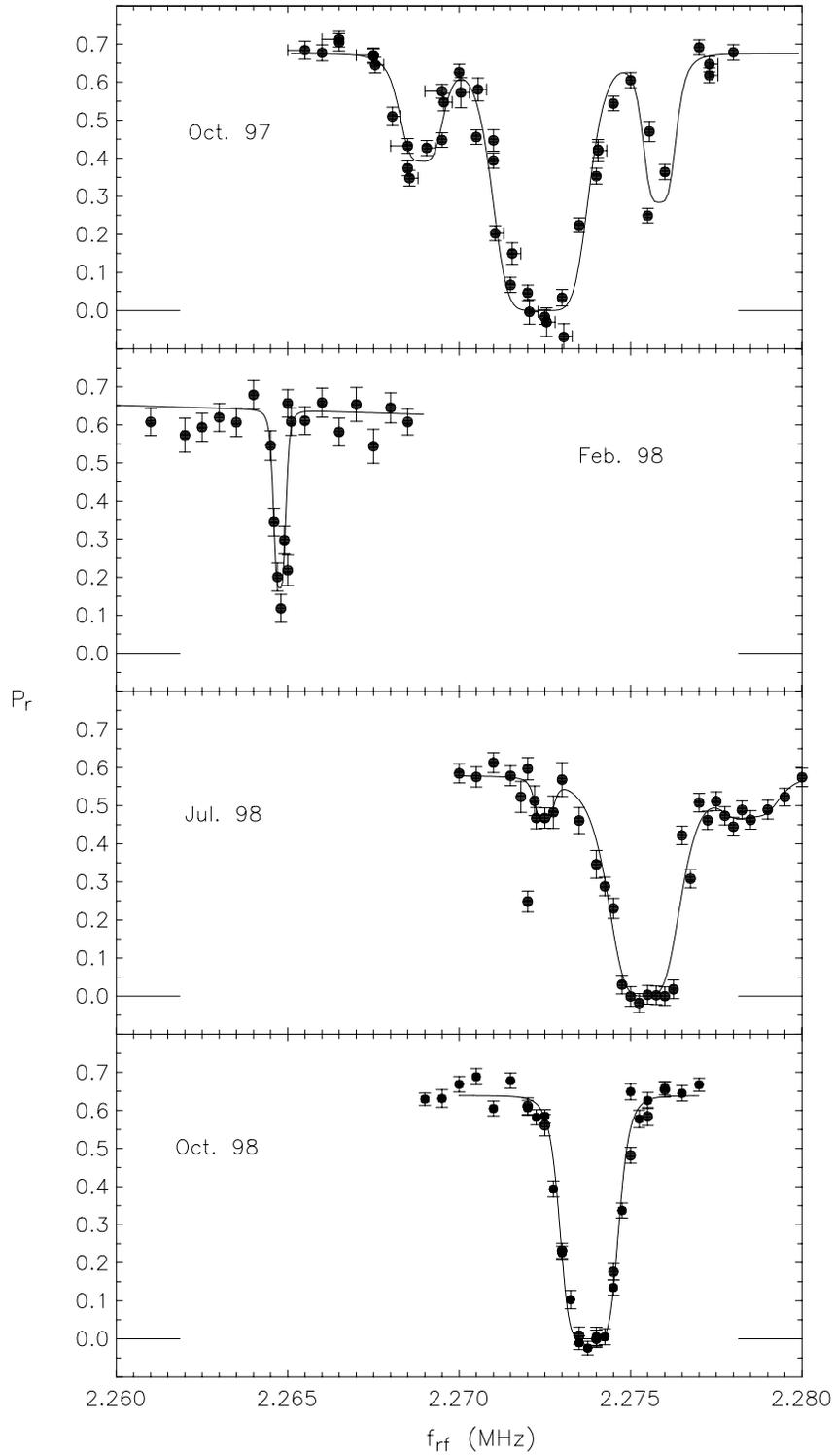}
\end{center}
\caption{\small RF solenoid depolarizing resonances. The measured radial beam polarization
at 104.1~MeV is plotted against the frequency of the rf solenoid.}
\end{figure}

In the July~1998 run we also used the rf dipole to create an rf depolarizing resonance; the
experimental data is shown in Figure~IV.2. 
\begin{figure} [h!]
\begin{center}
\epsfig{file=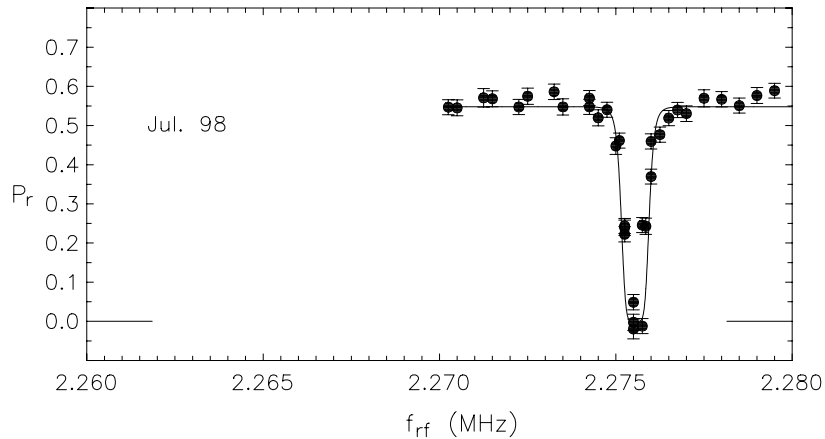}
\end{center}
\caption{\small RF dipole depolarizing resonance. The measured radial beam polarization
at 104.1~MeV is plotted against the frequency of the rf dipole.}
\end{figure}

We fit the rf depolarizing resonance's data in Figures~VI.1 and VI.2 using 
second-order and higher-order Lorentzians:
\begin{equation}
P=P_0\frac{(f-f_{r})^{2n}}{(f-f_{r})^{2n}+(w/2)^{2n}},
\end{equation}
where $n$ is the order of the Lorentzian, $P_0$ is the full polarization,
$f_{r}$ is the resonance's central frequency,
and $w$ is the resonance's FWHM width (full-width at half-maximum)). 

Note that two synchrotron sideband resonances were clearly present
in the October~1997 and July~1998 data~\cite{blinov3}; we 
fit them using higher-order Lorentzians with a variable depth:
\begin{equation}
P=P_0\left(d_{sync}\left(\frac{[f-(f_{r}\pm f_{sync})]^{2n}}{[f-(f_{r}\pm f_{sync})]^{2n}+(w_{sync}/2)^{2n}} - 1\right)+1\right),
\end{equation}
where $n$ is the order of the Lorentzian, $P_0$ is the initial polarization, $d_{sync}$ is the
synchrotron sideband's 
depth, $f_{r}$ is the main resonance's central frequency, determined by fitting the main dip,
$f_{sync}$ is the sideband's central frequency with respect to the main resonance's 
frequency, while the "+" and "-" signs represent the higher-frequency and lower-frequency 
sidebands, respectively, and $w_{sync}$ is the sideband's FWHM width. 
These parameters for each graph are summarized in Table~IV.2.
\begin{table}
\begin{center}
\begin{tabular}{|l|c|c|c|c|c|}
\hline
Run date & Oct. 97 & Feb. 98 & July 98 (sol.) & July 98 (dip.) & Oct. 98 \\
\hline
$f_{res}$ (MHz) & 2.27237 & 2.26477 & 2.27541 & 2.25549 & 2.27379 \\
$w$ (Hz) & 2890 & 387 & 2260 & 636 & 1810 \\
Lorentzian order & 3 & 1 & 2 & 3 & 3 \\
$f_{sync}^+, f_{sync}^-$ (Hz) & 3470, 3370 & - & 3075, 2924 & - & - \\
$w_{sync}^+, w_{sync}^-$ (Hz) & 1030, 1360 & - & 1840, 606 & - & - \\
$d_{sync}^+, d_{sync}^-$ & 0.45, 0.33 & - & 0.18, 0.19 & - & - \\
Sync. sideband &\raisebox{-0.08 in}{2, 2} & \raisebox{-0.08 in}{-}& \raisebox{-0.08 in}{2, 3} & \raisebox{-0.08 in}{-} & \raisebox{-0.08 in}{-} \\
\raisebox{0.35 in}{Lorentzian order} & & & & & \\
\hline
\end{tabular}
\end{center}
\caption{\small Locations and widths of rf depolarizing resonances at 104.1~MeV.}
\end{table}

Note that, in Figure~IV.1, the frequencies of some October~1997 data points were shifted by the 
amounts indicated by the single horizontal error bars. During the data taking, the snake solenoid was
twice refilled with liquid helium; the solenoid's current could not be exactly reproduced after each
fill, which resulted in a slightly different snake strength for each fill. 
This change in solenoid current of typically 0.02$\%$ caused a small change of the 
resonance's position for each fill; the frequencies of the data points are shifted
on the graph to account for this change.

\section{RF depolarizing resonance widths}
As mentioned in Section~II.4, the measured widths of these rf depolarizing resonances
was usually much larger than the simple single-particle 
model~\cite{widththeory} prediction. For an rf solenoid, this model predicts:
\begin{equation}
w_{th}^{sol} = f_c\epsilon^{sol} =f_c\frac{(1 + G)e\int{\!B_{rf}^{sol}\!\cdot\!dl}}{2\pi\gamma m_pv},
\end{equation}
where $f_c$ is the circulation frequency, $B_{rf}^{sol}$ is the r.m.s. rf magnetic field, while
$\gamma$ and $v$ are the proton's Lorentz factor and velocity respectively.
At the 6~kV maximum rf voltage across our rf solenoid, its 
$\int{\!B_{rf}^{sol}\!\cdot\!dl}$ was about 1.35~T$\cdot$mm; then Equation~IV.83
predicts an rf depolarizing resonance's width of about 600~Hz at 104.1~MeV.

However, as indicated in Table~IV.2 and Figure~IV.1, the measured resonances 
widths were usually about 2~kHz;
the only exception was the February~1998 run, when the measured width of 387~Hz was 
somewhat closer to the theoretical value.

For an rf dipole, a similar formula~\cite{widththeory} yields for the resonance width:
\begin{equation}
w_{th}^{d} = f_c\epsilon^{d} =f_c\frac{Ge\int{\!B_{rf}^{d}\!\cdot\!dl}}{2\pi m_pv}.
\end{equation}
At our 23~V maximum r.m.s. voltage across the dipole, its 
$\int{\!B_{rf}^{d}\!\cdot\!dl}$ was about 0.03~T$\cdot$mm, which gives
a predicted resonance
width of only about 110~Hz. As indicated in Table~IV.2 and Figure~IV.2, the 
width observed in July~1998 was about 1~kHz.

To determine possible sources of this width broadening, we recently measured resonance
widths at different conditions~\cite{width}. We found that the rf resonance's width depends
strongly on the rf acceleration voltage, which drives the synchrotron oscillations of the beam 
protons' energy; with these rf acceleration cavities turned on, the rf depolarizing resonance 
width was about
twice wider than with the rf cavities off. 
When the rf acceleration cavities were turned on at a high voltage of
about 1.5~kV, the energy of the stored protons oscillated with a 4~kHz synchrotron frequency 
around the equilibrium energy~\cite{sync}; the amplitude of these oscillations
depended on the beam's momentum spread $\Delta\!\!~p/p$, 
which was about $2\cdot10^{-4}$ with the electron cooling on.
Because of these synchrotron oscillations, each proton's circulation frequency $f_c$ and 
spin tune $\nu_s$ also oscillated with the same frequency of about 4~kHz and with amplitudes
$\Delta\!f_c/f_c$ and $\Delta\!\!~\nu_s$ of approximately 0.02$\%$
each. This spread in $f_c$ and $\nu_s$ resulted in some of the protons
having an $f_c$ and $\nu_s$
which satisfied the resonance
condition; this could partially depolarize the beam even when the rf solenoid or the
rf dipole was relatively far away from the resonance frequency, defined by Equation~IV.79 or IV.80.

We also found that the measured rf depolarizing resonance width
usually depended very weakly on the magnetic
field strength of either the rf solenoid or the rf dipole. When the strength of each rf magnet was
reduced by a factor of two, the measured rf resonance width only changed by about 5$\%$.
This indicates that the measured width of the rf depolarizing
resonance was not dominated by the resonance's strength.

\begin{figure} [h!]
\vspace{-0.25 in}
\begin{center}
\epsfig{file=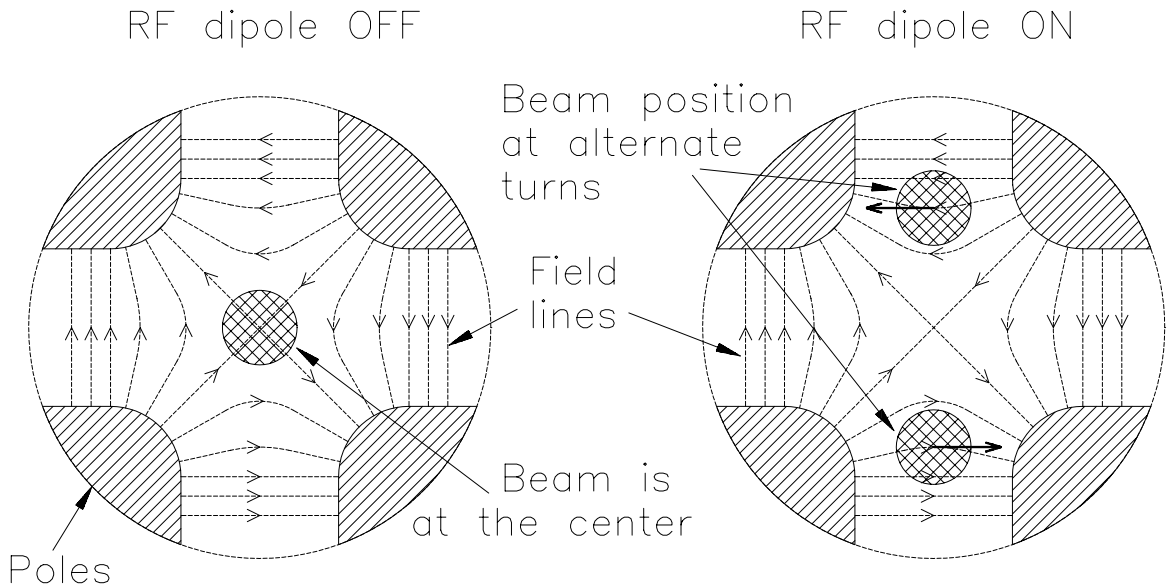}
\end{center}
\vspace{-0.15 in}
\caption{\small Possible rf dipole depolarizing resonance enhancement due to coherent betatron
oscillations in a quadrupole.}
\end{figure}
An interesting mechanism involving coherent betatron oscillations was suggested~\cite{dipreswidth} for the increase of an rf
dipole's resonance width. An rf dipole
with a horizontal magnetic field would increase the vertical betatron
oscillations' amplitude which could be large if the rf dipole's frequency is near the vertical betatron frequency.
Figure~IV.3 shows such coherent betatron oscillations when the
rf dipole frequency is a half-integer times the circulation frequency; the beam center's displacement
is then opposite at consecutive turns. 
The beam's protons
would all then encounter horizontal magnetic fields of this quadrupole, which are indicated by
solid arrows. In the protons'
rest frame, these fields' horizonal $\int\!B\!\cdot\!dl$ would have the rf dipole's frequency (1/2$f_c$) 
and thus could effectively increase the resonance's strength.
Such rf-dipole-induced coherent beam oscillations were successfully used to make a
Brookhaven AGS intrinsic
depolarizing resonance strong enough to spin-flip with no depolarization~\cite{bai}.

\section{Spin-flipping}
We spin-flipped a stored horizontally-polarized proton beam using either
the rf solenoid or the rf dipole by sweeping the magnets' frequency through the rf
depolarizing resonance found earlier in the experiment. The main goal
of our spin-flipping studies was to maximize the spin-flip efficiency $\eta$, 
defined by Equation~II.70. 

To maximize the efficiency of the spin-flip we varied the frequency ramp time 
$\Delta~\!t$ and the frequency ramp range $\Delta\!f$, while measuring the
beam polarization after a spin-flip; these two parameters
determine the resonance crossing rate $\Delta\!f/\Delta~\!t$. According
to the Froissart-Stora formula (Equation~I.7), a slower rate of crossing the
resonance yields a higher final polarization, and thus a higher
spin-flipping efficiency $\eta$.
To make the crossing rate $\Delta\!f/\Delta~\!t$ slower, one can either increase
the ramp time $\Delta~\!t$ or reduce the frequency range $\Delta\!f$.

Increasing the ramp time while holding $\Delta\!f$ fixed
is technically quite easy; this suggests that a very long $\Delta~\!t$ should allow 
an almost perfect spin-flip.
However, our earlier experiments~\cite{caussyn} found
that at very long ramp times (greater than about 500~ms) this efficiency decreases.
Thus, there may be an optimum setting for $\Delta~\!t$, which gives the highest
final beam polarization.

Varying the frequency range $\Delta\!f$, while holding $\Delta~\!t$ fixed,
can change the spin-flip efficiency in two ways: by changing the crossing rate 
$\Delta\!f/\Delta~\!t$ and by making the resonance crossing incomplete 
at narrow frequency ranges.
The first mechanism dominates when $\Delta\!f$ is much wider than the resonance width $w$;
thus, to improve the spin-flipping efficiency one should decrease $\Delta\!f$,
which reduces the resonance crossing rate. However, when the frequency range becomes
smaller than the width of the resonance, the second mechanism becomes important. 
The rf magnet is then turned on and off
at frequencies where the beam is already partially depolarized; this certainly reduces 
the final polarization after a spin-flip. Clearly, one should choose the optimum 
$\Delta~\!t$ and $\Delta\!f$ to simultaneously minimize
both of these efficiency-reducing mechanisms.

\subsection{Experimental procedure}
During the spin-flipping studies, the experimental procedure differed slightly
from the procedure used to map an rf resonance. After injecting, storing and
cooling the beam, the Wavetek waveform generator (Figure~III.8) 
was triggered and then the rf magnet was turned on; 
following the control signal from the Wavetek, the rf magnet then ran for about
10~ms at a frequency $f_r-\Delta\!f/2$. During
a time $\Delta\!~t$, 
the Wavetek ramp then linearly increased the rf magnet's frequency to
a frequency above the resonance's central frequency $f_r$; 
the Wavetek then kept the rf magnet at this frequency $f_r+\Delta\!f/2$ 
for the rest of the flattop.

\subsection{Spin-flipping by varying the ramp time}
The location and width of the rf depolarizing resonance used
for spin-flipping were first determined from a resonance mapping 
such as in Figure~IV.1 and IV.2. We then set the frequency range $\Delta\!f$ so that
it completely bracketed the rf resonance by extending symmetrically
on both sides by about two resonance widths; then we varied the ramp time $\Delta~\!t$. 
\begin{figure}
\vspace{-0.2 in}
\begin{center}
\epsfig{file=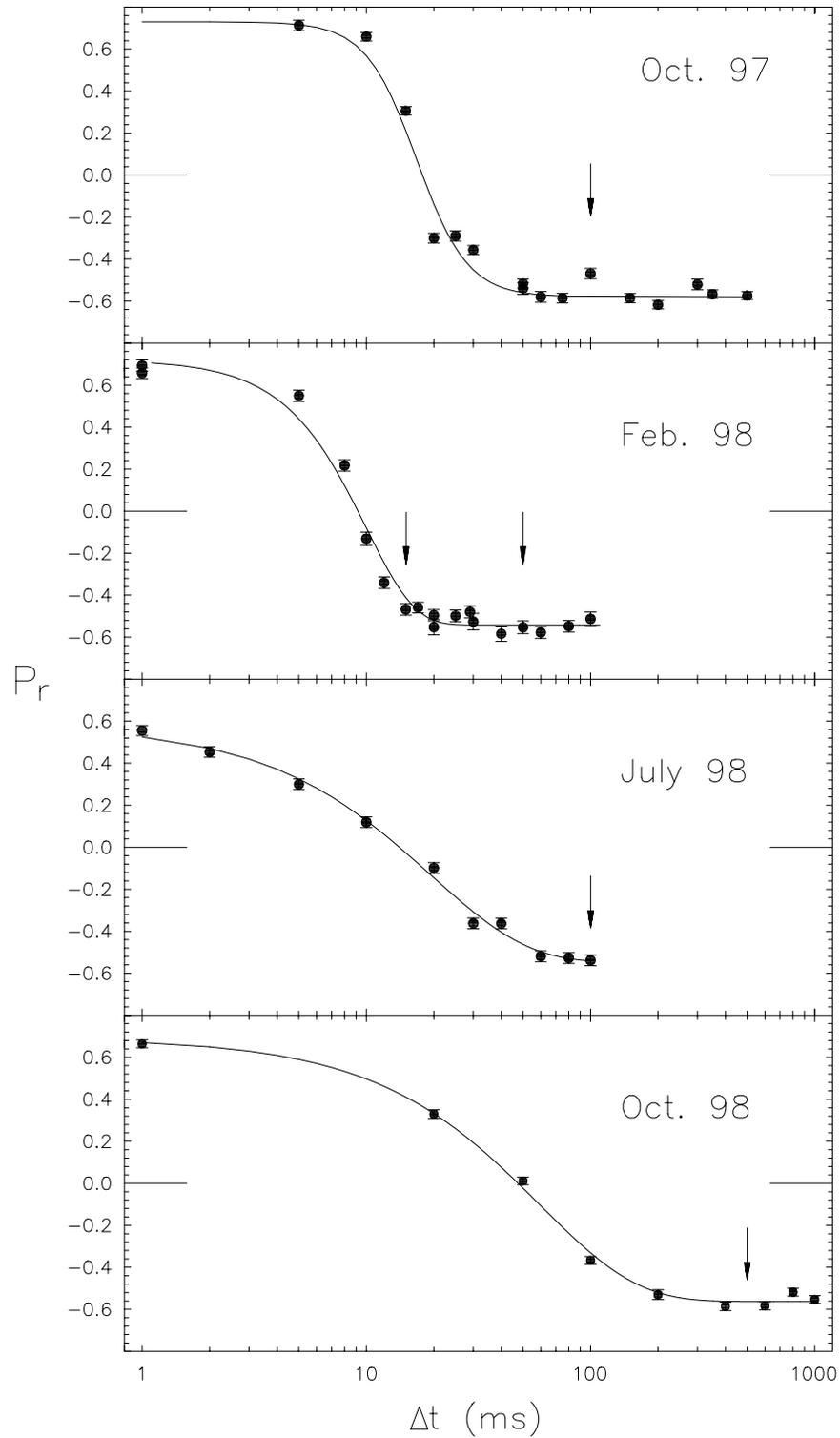}
\vspace{-0.4 in}
\end{center}
\caption{\small Spin-flipping while varying the rf solenoid's ramp time. The measured radial beam
polarization at 104.1~MeV is plotted against the ramp time $\Delta\!~t$. Arrows indicate
the ramp times, used in spin-flipping while varying the frequency range, as described in
Section~IV.4.3.}
\end{figure}

Figure~IV.4 shows the results of spin-flipping using the rf solenoid 
by varying its ramp time; the measured radial
beam polarization after a single spin-flip is plotted against the ramp time
$\Delta~\!t$. With the exception of the February~1998 data, the data points did not
follow the Froissart-Stora prediction, which was shown in Figure~II.3. 
We fit the July~1998 and October~1998 graphs data using the empirical equation:
\begin{equation}
P_f = P_i[(1+\eta_M)e^{- \frac{(\pi \epsilon f_c)^2}{(\Delta f / \Delta t)}}-\eta_M],
\end{equation}
where $\eta_M$ is the maximum spin-flip efficiency, 
and $\epsilon\!f_c$ is the normalized resonance strength. 
This equation is a modified Froissart-Stora formula, incorporating an 
imperfect maximum spin-flip efficiency
parameter $\eta_M$;
recall that the Froissart-Stora formula (Equation~I.7) predicts a 100$\%$ efficiency 
at very long ramp times.

To fit the October~1997 and February~1998 data, this empirical formula had to be even
further modified. In these runs, a flat region was observed at very short ramp times of up to about
10~ms. To fit this flat region, we empirically raise the exponent in Equation~IV.85 to the second 
power to yield:
\begin{equation}
P_f = P_i[(1+\eta_M)e^{- \left(\frac{(\pi \epsilon f_c)^2}{(\Delta f / \Delta t)}\right)^2}-\eta_M].
\end{equation}

The fit parameters in both cases include: the initial beam polarization $P_i$, the 
maximum spin-flip efficiency $\eta_M$ and the normalized resonance 
strength $\epsilon\!f_c$; these parameters are listed in Table~IV.3.
\begin{table} [b!]
\begin{center}
\begin{tabular}{|l|c|c|c|c|}
\hline
Run date & Oct. 97 & Feb. 98 & July 98 & Oct. 98 \\
\hline
$P_i$ & 0.738 & 0.721 & 0.582 & 0.691 \\
$\eta_M$ & 0.720 & 0.750 & 0.940 & 0.815 \\
$\epsilon\!f_c$ (Hz) & 513 & 283 & 500 & 286 \\
\hline
\end{tabular}
\end{center}
\caption{\small Fit parameters for spin-flipping by varying the ramp time at 104.1~MeV.}
\end{table}

Note that for the October~1997, July~1998 and October~1998 data, the resonance strength, 
measured by spin-flipping, was much smaller than the mapped resonance width; in fact 
the spin-flipping resonance strength
\begin{figure}[b!]
\vspace{-0.3 in}
\begin{center}
\epsfig{file=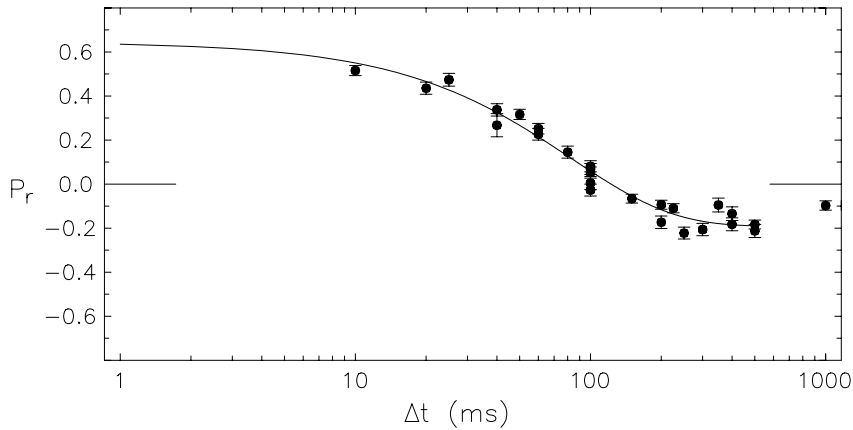}
\end{center}
\caption{\small Spin-flipping while varying the rf dipole's ramp time. The measured radial beam
polarization at 104.1~MeV is plotted against the ramp time $\Delta\!~t$.}
\end{figure}
was quite close to the predicted value of about 600~Hz. For the February~1998 data, both the mapped
resonance width and the resonance strength measured by spin-flipping agree very well.
This may be another indication that the apparent resonance width increase is due 
to some rf structure of the beam and not a manifestation of the rf depolarizing resonance's
actual strength.

We also attempted to spin-flip a horizontally polarized beam using the vertical-field rf
dipole described in Chapter~III. The radial beam polarization after a single spin-flip
is plotted against the rf dipole's ramp time in Figure~IV.5. Note that the maximum beam
polarization after spin-flip was only about 23$\%$, while the initial polarization was 60$\%$;
this indicates a spin-flipping efficiency of only about 38$\%$. Note also that, up to about
100~ms, the final polarization direction is not yet reversed; this suggests that the
rf dipole-induced depolarizing resonance is very weak. Fitting the data using Equation~IV.85,
while ignoring the data point at 1000~ms, we get a normalized resonance strength
$\epsilon\!f_c$ of 240~Hz. As in the case of the rf solenoid-induced depolarizing
resonance, this normalized strength is much smaller than the measured width $w$,
and agrees better with the prediction given by Equation~IV.84.

Probably, the low rf-dipole spin-flip efficiency is due to the rf dipole's weak
field. We plan to install a variable capacitor in parallel with the rf dipole to
form an LC resonant circuit, similar to the rf solenoid's. This resonant circuit may
help to increase the voltage across the rf dipole, and thus its magnetic field and
resonance strength.

\subsection{Spin-flipping by varying the frequency range}

After determining the optimum ramp time, we set $\Delta~\!t$ to that
optimum value, which was, for example, 100~ms in
October~1997, as indicated in Figure~IV.4. 
We then studied spin-flipping while varying the frequency range
to determine its optimal setting.
Figure~IV.6 shows the measured radial beam polarization plotted against the frequency range
$\Delta\!f$ at a fixed ramp time; the ramp time for each curve is
shown on the graph. 

Most of the $\Delta\!f$ curves have a rather pronounced maximum where the
spin-flipping efficiency is the highest. The very narrow maximum in the 15~ms data from
February~1998 indicates that this 15~ms ramp time was too short; thus even a small increase
in the resonance crossing rate $\Delta\!f/\Delta~\!t$ due to a frequency range increase, 
significantly reduced the spin-flip efficiency; when the ramp time
was increased to 50~ms, the maximum became much broader. For practical applications
of spin-flipping in high-energy scattering experiments, it would be
desirable to have some freedom in choosing
the spin-flipper parameters; longer ramp times
allow more freedom with the $\Delta\!f$ choice. However, as was
mentioned in Section~VI.4, the spin-flipping efficiency seems to drop at very long ramp times.
\begin{figure}
\begin{center}
\epsfig{file=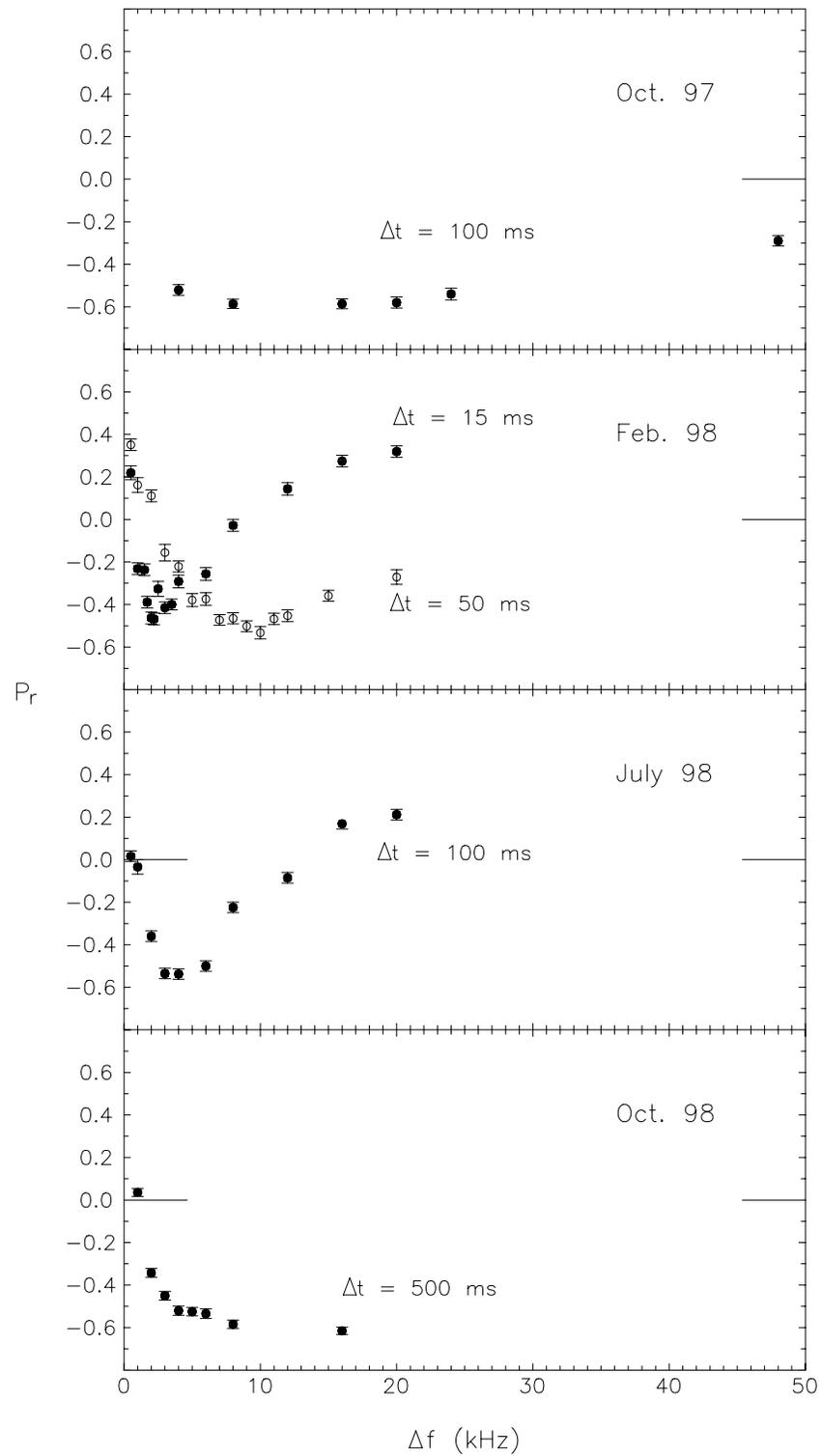}
\end{center}
\caption{\small Spin-flipping by varying the rf solenoid's frequency range. 
The measured radial beam polarization at 104.1~MeV is plotted against the 
frequency range $\Delta\!f$.}
\end{figure}

Note that both the October~1997 and July~1998 data were taken with a ramp time $\Delta\!~t$
of 100~ms. However, in October~1997 the spin-flip polarization remained at a high value of 
about 60$\%$ for 
$\Delta\!f$ of up to about 25~kHz, while in July~1998, it remained large only up to
about 7~kHz and then decreased to zero near 15~kHz and then changed sign. 
Since the resonance crossing rate
$\Delta\!f/\Delta~\!t$ was the same for both data sets, the data may indicate
that the actual resonance strength was probably much larger in October~1997
than in July~1998; this allowed more efficient spin-flipping even at faster ramps.
Another possible explanation is that the synchrotron sidebands were present around
the main resonance in July~1998, which could reduce the spin-flipping efficiency.

In the October~1998 data, we saw
the polarization decrease only at small values of $\Delta\!f$, while it was fairly flat in the 
frequency ramp range of 4 to 16~kHz; this was probably due to a very long ramp time of 500~ms.
The resonance crossing rate $\Delta\!f/\Delta~\!t$ was then sufficiently slow for a nearly complete 
spin-flip even at high $\Delta\!f$. Moreover,
the polarization decrease at values of $\Delta\!f$ below about 4~kHz resembles the shape
of the rf depolarizing resonances, discussed in Section~IV.2. 

Recall that, in Figure~IV.6, a
frequency range of $\Delta\!f$ represents varying the rf solenoid's
frequency from $f_r-\Delta\!f/2$ to $f_r+\Delta\!f/2$. The HWHM (half-width
at half-maximum) is 
the point on the curve where the polarization is
1/2 of the maximum polarization; this HWHM $\Delta\!f$ width should be equal to the 
measured width of the resonance.
Indeed, in the July~1998 and October~1998 curves, the HWHM $\Delta\!f$ defined this way was about
2~kHz, which is very close to the measured widths of 2260~Hz and 1810~Hz, 
shown in Table~VI.2.
The February~1998 data with $\Delta\!~t=15$~ms has a HWHM in $\Delta\!f$ of approximately
1~kHz, while the February~1998 data with $\Delta\!~t=50$~ms has a HWHM in 
$\Delta\!f$ of about 4~kHz; both are very
different from the 387~Hz, shown in Table~VI.2. However, the polarization is positive at very
low $\Delta\!f$. This may indicate that the resonance frequency $f_r$ for Figure~VI.6 may have shifted
slightly by an amount $\delta\!f_r$ from the measured position when it was mapped; 
the frequency ramp then may have started not exactly
at $f_r-\Delta\!f/2$, but closer or further away from the resonance. For $\Delta\!f/2~=~\delta\!f_r$,
the frequency sweep would either start or end exactly on the resonance; thus one could then 
expect complete a depolarization of the beam. For $\Delta\!f/2<\delta\!f_r$ the resonance would not
be crossed, so there would only be partial depolarization without spin-flip.

\subsection{Multiple spin-flipping}
With the ramp time and frequency range set to maximize the spin-flip efficiency,
we performed multiple spin-flipping with the rf solenoid by crossing the resonance several times 
and measuring the beam's polarization after these multiple crossings. This
technique was used to obtain a much more precise measurement of the spin-flip efficiency.
\begin{figure} [b!]
\begin{center}
\epsfig{file=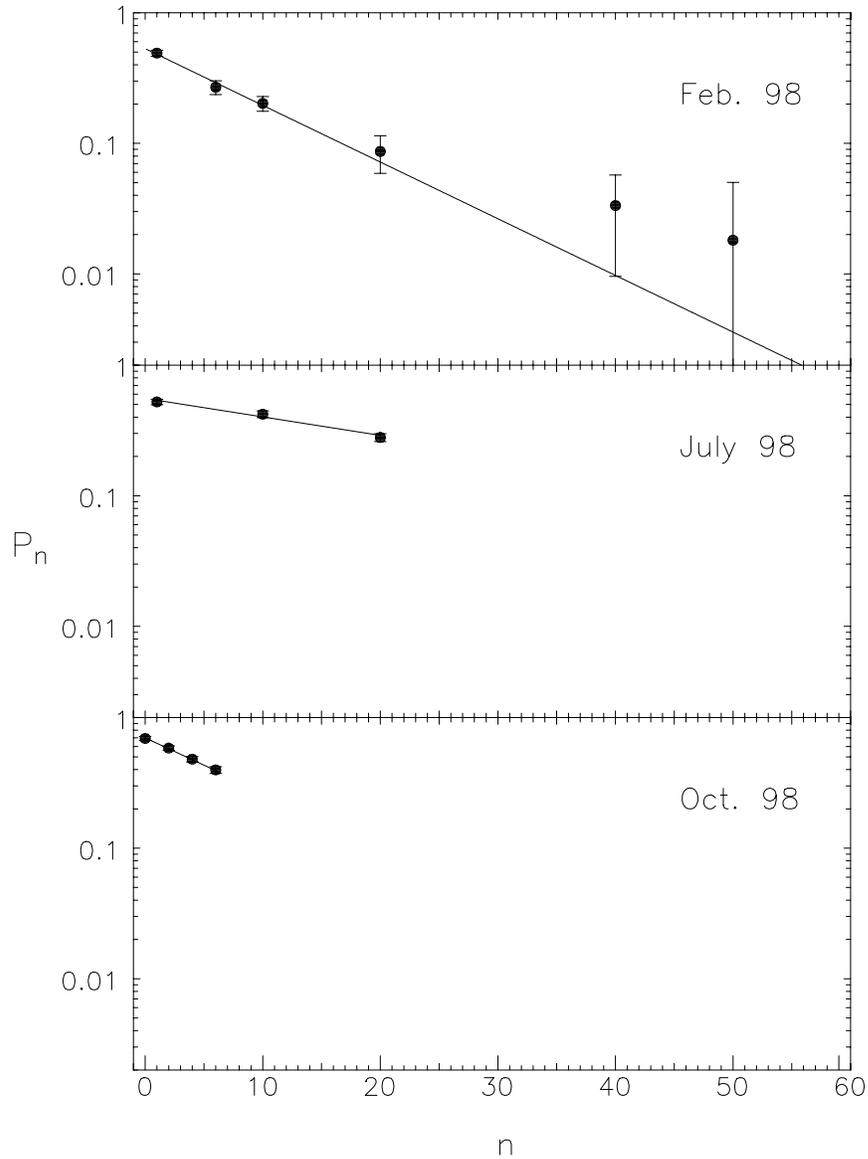}
\end{center}
\caption{\small Multiple spin-flipping using the rf solenoid.The measured radial beam 
polarization at 104.1~MeV is plotted against the number of spin-flips n.}
\end{figure}

We did not perform multiple spin-flipping in the October~1997 run due to lack of beam time. 
However, the multiple spin-flip data for the February~1998, July~1998 and October~1998 
runs are shown in Figure~IV.7; in each graph, the
radial beam polarization is plotted against the number of spin-flips.
We always flipped the beam protons' spins an even number of times, so that the polarization direction
after multiple spin-flipping was the same as the initial polarization direction. 
The magnitude of the final polarization was generally reduced by some depolarization 
during each spin-flip. We fit the multiple spin-flip data using
\begin{equation}
P_n = P_i\eta^n,
\end{equation}
$P_n$ is the beam's polarization after $n$ spin-flips, $P_i$ is the beam's initial polarization, 
and $\eta$ is the efficiency of a single spin-flip. The maximum spin-flip efficiencies achieved 
in each experimental run are summarized in Table~IV.4. Note that the efficiencies measured
in multiple spin-flipping are significantly higher than those measured by a single spin-flip. 
\begin{table}[h!]
\begin{center}
\begin{tabular}{|l|c|c|c|c|}
\hline
Run date & Oct. 97 & Feb. 98 & July 98 & Oct. 98 \\
\hline
$\eta_M$ (single flip) & 0.720$\pm 0.025$ & 0.750$\pm 0.070$ & 0.940$\pm 0.050$ & 0.815$\pm 0.025$ \\
$\eta$ (multiple flips) & - & 0.910$\pm 0.010$ & 0.987$\pm 0.004$ & 0.950$\pm 0.010$ \\
\hline
\end{tabular}
\end{center}
\caption{\small Summary of spin-flipping efficiencies}
\end{table}

Indeed, the observed polarization loss in the first spin-flip is generally higher than
in the following flips. For example, in the February~1998 run, the polarization dropped by
about 25$\%$ in the first spin-flip, while in each following spin-flip the loss was only about
9$\%$. This higher polarization loss in the first spin-flip may be due to improper alignment 
of the stable spin direction inside the rf 
solenoid with respect to the rf solenoid's longitudinal magnetic field. With a full longitudinal
Siberian snake in the S section of the Cooler Ring, as was shown in Figure~III.4, 
the stable spin direction (SSD) at 104.1~MeV was at an angle of approximately $60^{\circ}$
to the rf solenoid's longitudinal field in the G section. The best 
spin-flipping results should occur when the SSD is perpendicular to the rf field direction.
If the SSD makes an angle $90^{\circ}-\alpha$ with respect
to the rf field, then the spin-polarization vector's
rotation around the rf field could smear out the polarization,
so that the polarization $P_f$ after a single flip would be reduced by $\cos{\alpha}$.
Since $\alpha$ is about $30^{\circ}$ at 104.1~MeV,
the beam polarization after one spin-flip would then be reduced by
$\cos{30^{\circ}}\simeq0.87$ thus limiting the single spin-flip efficiency by this value. 
In further spin-flips this effect apparently does not play a role.
Thus, total efficiency of the first spin-flip is a product of $\cos{\alpha}$ and the 
spin-flip efficiency measured using multiple flips. 

For the February~1998 and October~1998
data, this hypothesis produces rather good agreement with the data, giving single
spin-flip efficiencies of 0.79 and 0.82, respectively, using $\alpha=30^{\circ}$ and the 
measured multiple spin-flip efficiencies. However, for the July~1998 data the
calculated single spin-flip efficiency was about 0.85, while it was measured to be 0.943.
It is likely that there are some subtleties in the spin-flipping process, which are not yet 
understood.

\chapter{Conclusions}

Our 1997-1998 experiments at the IUCF Cooler Ring show that it is possible to 
use a longitudinal rf magnet (rf solenoid) to spin-flip a horizontally
polarized proton beam in the presence of a full Siberian snake.
In July 1998, we reached a maximum spin-flipping efficiency of 98.7$\pm$1$\%$
in multiple spin-flips
by adjusting the rf solenoid's frequency ramp time and frequency range. 
This efficiency corresponds to about a 1.3$\%$ polarization loss per spin-flip;
with an initial beam polarization of 70$\%$, this leads to a  
polarization of about 36$\%$ after 50 flips.

However, due to the Lorentz contraction of an rf solenoid's longitudinal magnetic
field integral, the strength of an rf-solenoid-induced
depolarizing resonance decreases at higher beam energies, which makes an rf solenoid impractical
for spin-flipping in high energy storage rings such as RHIC and HERA. 
An rf-dipole magnet, whose transverse
$\int\!B\!\cdot\!dl$ is independent of the beam energy
seems more suitable for efficient spin-flipping at high energy. 

Studies of rf-dipole spin-flipping a vertically polarized stored proton beam 
at the IUCF Cooler Ring with no Siberian snake present
were quite successful; a spin-flip efficiency of $96.7\pm\!1.0\%$ was
achieved~\cite{dipflipnosnake}. However, in the presence of a nearly 
full Siberian snake, spin-flipping 
a horizontally polarized proton beam with an rf-dipole 
has so far had little success; the maximum spin-flip efficiency was below 40$\%$.
The main reason for this low spin-flip efficiency may be the weaker strength
of our vertical rf dipole, whose $\int\!B\!\cdot\!dl$ was only about 2$\%$ of the
rf solenoid's.
We are currently working on icreasing the rf-dipole 
spin-flipping efficiency for a horizontally polarized beam; we plan to try increasing 
the existing rf dipole's strength by using an LC resonant circuit, and way later build
a new, higher-field rf dipole.

We also found that the measured widths of depolarizing resonances, created
by both the rf solenoid and the rf dipole, were often much larger than the width predicted by
the measured resonance strength, which
was obtained from the spin-flipping data, and agrees quite well 
with theoretical predictions.
We recently studied this relation between width and strength at the IUCF Cooler Ring~\cite{width};
a preliminary conclusion of these studies is that the apparent widening of the
resonance may be due to the complex rf structure of the beam. Reducing the
rf solenoid's or rf dipole's $\int\!B\!\cdot\!dl$ by a factor of 2 did not significantly
reduce the measured resonance's width. However, turning
off the rf acceleration cavities, which may generate this beam structure, reduced
the measured resonance width by a factor of about 2.

\startappendix
\begin{theappendix}

\startanappendix{CE-69 COLLABORATION} \label{coll}

\setsinglespacing
\noindent{\bf University of Michigan}: 
V.A. Anferov$^1$, B.B. Blinov, M.A. Bychkov$^2$, E.D.~Courant$^3$,
Ya.S.~Derbenev, A.D.~Krisch, W.B.~Lorenzon, W.A.~Peters,
R.A.~Phelps$^4$, L.G.~Ratner$^{3\dagger}$, K.V.~Sourkont$^2$, V.K.~Wong

\setsinglespacing \vspace{1pc}
\noindent
{\bf Indiana University Cyclotron Facility}: C.M.~Chu,
S.Y.~Lee, T.~Rinckel, P.~Schwandt, B.~von~Przewoski

\setdoublespacing

\noindent
{\bf HEARO (INS and KEK), Japan}: C. Ohmori, H. Sato, T. Toyama

\noindent
{\bf Fermilab}: P.S. Martin, A.D. Russell

\noindent 
{\bf King Fahd University}: F.Z. Khiari

\noindent 
{\bf SLAC}: A.W. Chao, M.G. Minty

\setsinglespacing\vspace{1pc}\noindent $^1$ now at IUCF

\noindent $^2$ also at Moscow State University

\noindent $^3$ also at Brookhaven National Laboratory 

\noindent $^4$ now at IBM, Inc. 

\noindent $^{\dagger}$ deceased

\vspace{1pc}\noindent
Note:  The institutions listed are the institutions where each person
was primarily 
working during the CE-69 experiment (1997-1998).  Many people have
since moved.

\end{theappendix}

\startbibliography

\end{document}